\documentclass[proof]{pasj00}
\draft

\usepackage{lscape}
\begin{document}
\SetRunningHead{Author(s) in page-head}{Running Head}
\Received{}
\Accepted{}

\title{Chemical abundances of the outer halo stars in the Milky Way \thanks{Based on data collected at the Subaru Telescope, which is operated by the National Astronomical Observatory of Japan}}

\author{Miho \textsc{Ishigaki}}
\affil{Astronomical Institute, Tohoku University, Sendai 980-8578, Japan}
\email{miho@astr.tohoku.ac.jp}

\author{Masashi \textsc{Chiba}}
\affil{Astronomical Institute, Tohoku University, Sendai 980-8578, Japan}\email{chiba@astr.tohoku.ac.jp}
\and
\author{Wako {\sc Aoki}}
\affil{National Astronomical Observatory, Mitaka, Tokyo 181-8588, Japan}\email{aoki.wako@nao.ac.jp}


%


\maketitle

\begin{abstract}
We present chemical abundances of $57$ metal-poor ([Fe/H]$<-1$) 
stars that are likely constituents of the outer stellar halo in the Milky Way. 
Almost all of the sample stars have an orbit reaching a maximum vertical distance
($Z_{\rm max}$) of $>5$ kpc above and below the Galactic plane. 
High-resolution ($R\sim 50000-55000$), high signal-to-noise (S/N$>100$)
spectra for the sample stars obtained with Subaru/HDS are 
used to derive chemical abundances of Na, Mg, Ca, Ti, Cr, Mn, Fe, 
Ni, Zn, Y and Ba with an LTE abundance analysis code. 
The resulting abundance data are combined with those presented in literature 
that mostly targeted at smaller $Z_{\rm max}$ stars, 
and both data are used to investigate any systematic trends 
in detailed abundance patterns 
depending on their kinematics. It is shown that, 
in the metallicity range of $-2<$[Fe/H]$<-1$, the [Mg/Fe] 
ratios for the stars with $Z_{\rm max}>5$ kpc are systematically 
lower ($\sim$0.1 dex) than those with smaller $Z_{\rm max}$. 
For this metallicity range, a modest degree of 
depression in the [Si/Fe] and the [Ca/Fe] ratios are also observed. 
This result of the lower [$\alpha$/Fe] for the assumed outer halo 
stars is consistent with previous studies that found a signature of 
lower [$\alpha$/Fe]
 ratios for stars with extreme kinematics. 
A distribution of the [Mg/Fe] ratios for the outer halo stars 
partly overlaps with that for 
stars belonging to the Milky Way dwarf satellites in the metallicity 
interval of $-2<$[Fe/H]$<-1$ and spans a 
range intermediate between the distributions for the inner halo stars  
and the stars belonging to the satellites.
Our results confirm inhomogeneous nature of chemical abundances within the 
Milky Way stellar halo depending on kinematic properties of constituent 
stars as suggested by earlier studies. 
Possible implications for the formation of the Milky Way halo 
and its relevance to the suggested dual nature of the halo are discussed.

\end{abstract}

\section{Introduction}

Based on recent theoretical and observational studies, 
it is now widely believed that 
the stellar halo of the Milky Way (MW) Galaxy has formed 
through a series of accretions of smaller subsystems  
in the course of the hierarchical galaxy formation in the Universe. 
Recent numerical simulations, based on the standard 
$\Lambda$CDM model, predict that such accretion events 
leave various substructures within a stellar halo of 
MW-sized ($\sim10^{12} \MO$) galaxies (e.g., \cite{key-59}). 
This prediction is well supported by discoveries of various 
substructures in the MW halo (e.g., 
\cite{key-60}) as well as in the halos of nearby galaxies 
(e.g., \cite{key-61, key-62, key-63}) as can be identified 
in spatial distribution, kinematics and metallicity 
of constituent stars. 
As an example, recent extensive photometric surveys of the 
MW halo reveal 
clumpy spatial distributions of stars being 
apart from the smooth underlying component, such as 
Sagittarius dwarf galaxy \citep{key-18,key-21} or 
Virgo overdensity \citep{key-23,key-91}. 
Furthermore, a kinematic
 substructure of stellar tidal debris remains longer than a spatial 
substructure, providing insights into 
 the earlier stage of the MW formation 
(e.g., \cite{key-66, key-22}). Indeed, presence of substructures in 
a velocity or an angular-momentum space has been 
reported by several authors 
(e.g., \cite{key-26, key-4, key-27, key-52, key-53, key-56, key-65, key-89}).

Further observational constraints on the formation of the MW 
stellar halo have been suggested from the fact that the halo 
comprises of two structural components, the inner and the outer halos, 
using kinematics and metallicity (e.g., \cite{key-81, key-80, key-4, 
key-35, key-90}). In particular, \citet{key-35} showed 
using SDSS data that the inner halo exhibits a modest net 
prograde rotation and the average metallicity of [Fe/H]$\sim-1.6$
\footnote{[Fe/H]$=\log A{\rm (Fe)}-\log A_{\odot}{\rm (Fe)}$}
while the outer halo has a net retrograde rotation with 
the lower average metallicity ([Fe/H]$\sim-2.2$). 
Based on these results, they 
suggested that the inner and the outer halos cannot be formed 
at once through a monolithic collapse but may have been formed through different 
mechanisms with different timescales and formation epoch.  
In particular, the results are consistent with the hypothesis 
that the accretion of dwarf spheroidal galaxies (dSphs) have played more 
important role in the outer halo. 

These recent discoveries of substructures as well as characteristic 
features of the inner and the outer halos provide us with 
a concrete picture for the current state of the MW stellar halo as 
being a collective entity of accreted subsystems and their 
tidal debris (e.g., \cite{key-64}). However, the 
origins of individual structures have not yet been well constrained
through phase space information alone because a relaxation process 
during the dynamical evolution of the MW progressively 
smears out such substructures and because the accurate 
measurements of distances and kinematics are yet limited. 

It has been recognized that detailed chemical abundance of 
metal-poor stars provides a 
fossil signature of the early epoch of the MW formation 
(e.g., \cite{key-67}).  Important advantage of using chemical 
abundance of metal-poor stars as a tracer is that the metal-poor 
stars formed early in the Universe and thus retain materials 
from that epoch in their atmosphere. Therefore, abundance 
pattern of the stellar atmosphere reflects that of the 
star forming clouds at the formation sites. In 
particular, ratios of the abundance of key elements 
including $\alpha$-elements (Mg, Si, Ca and Ti), 
Fe-peak elements (Cr, Ni, etc.) and neutron capture (Y, Ba, etc.) elements can be 
used to infer star formation history based on chemical evolution 
models, where these elements are predominantly ejected 
through either Type II or Type Ia SNe. 

Although the nucleosynthesis 
mechanisms of these elements are not fully understood, they 
provide unique constraints on the halo formation especially 
when they are used together with kinematics.
Indeed, important insights about the early evolution of 
the MW have been obtained through the analysis of chemical 
abundance patterns of field halo stars in combination with their 
kinematics (e.g., \cite{key-68, key-69, key-54, key-8}; 
\cite{key-7}, hereafter SB02; 
\cite{key-9}, \cite{key-15, key-36}). 
SB02 suggested that $\alpha$-elements 
to iron abundance ratios ([$\alpha$/Fe]) 
weakly decrease with increasing apogalactic distance 
along the galactic plane ($R_{\rm apo}$) 
for their $\sim 60$ sample of metal-poor ($-4<$[Fe/H]$<-1$) halo stars. 
\citet{key-8} also reported the weak correlation of abundance 
ratios for $\alpha$ elements with the Galactic rest frame velocity, 
$V_{\rm RF}$. 
Both of their sample of field halo stars were reported to have 
distinct abundances compared to the MW dwarf satellite galaxies.
 These results are inconsistent with a crude expectation that the MW halo 
stars may exhibit similar abundance to the MW satellites if dominant 
halo populations have been accreted from systems similar to these satellites. 

A possible interpretation for the abundance discrepancy between the 
field halo and the MW satellites has been  
investigated with numerical simulations that implement both cosmologically 
motivated mass accretion history and chemical evolution for individual 
building blocks (\cite{key-19, key-20}, \yearcite{key-55}; \cite{key-58}). 
\citet{key-20} suggest that the discrepancy in the [$\alpha$/Fe] ratios 
is a natural expectation from the standard $\Lambda$CDM model if 
the majority of the MW halo 
was formed by accretions of a few massive ($\sim 10^{8}-10^{10} \MO$) 
subhalos that had been disrupted early on
($> 8-9$ Gyr ago). In their model, the proposed progenitor subhalos 
(``building blocks'' of the MW host halo) 
had been enriched with $\alpha$ elements ejected 
from Type II SNe associated with a rapid formation of massive stars. 
The model assumes that the rapid enrichment with the 
Type II SNe ejecta was subsequently truncated due to the accretion 
onto the MW halo which can result in enhanced
 [$\alpha$/Fe] ratios commonly observed among the present-day halo stars. 
 In contrast, the surviving satellite galaxies are modeled to be enriched with iron-rich 
materials ejected with Type Ia SNe, whose explosion rate 
is expected to peak at a later time than Type II SNe 
as modeled by \citet{key-71}, resulting in the observed 
lower [$\alpha$/Fe] ratios.

Following these theoretical implications, 
it is now important to measure abundances of $\alpha$ and 
other key elements for individual stars over a large volume 
of the stellar halo
in order to obtain further constraints on properties and 
an assembly history 
of the possible building blocks.
In particular, in spite of the fact that 
the outer halo exhibits characteristics of 
hierarchically accreted stellar system 
in contrast to the inner halo \citep{key-35}, 
abundance data for the outer halo stars are 
currently quite insufficient.
In obtaining high-resolution spectra 
with required quality and statistics for 
the halo stars, however, a fundamental problem 
is that the survey volume is limited to the solar neighborhood 
(within $\lesssim 2$ kpc). Therefore, the kinematic information 
of stars is essential to select stars in remote locations of the halo.
The kinematic information including radial velocities, 
proper motions and distant estimates from the Hipparcos mission 
and other grand-based surveys are available for 
calculating orbital parameters of such halo stars \citep{key-5, key-4}.
These data can be used to select targets for the 
candidate outer halo stars for the purpose of 
examining similarities and differences 
in detailed chemical abundances among the recently reported inner 
and outer halo populations \citep{key-35, key-36}.  

In this paper, we analyze the dependence of chemical abundance 
of halo stars on their 
kinematics, based on a large data set assembled from 
our observations and literature. 
The most significant 
point of this work is the inclusion of a homogeneous sample of 57 
metal-poor stars whose orbits reach a maximum vertical distance 
from the galactic plane ($Z_{\rm max}$) greater than 5 kpc. 
Our initial results using a subset of 26 stars observed in 2003 
were presented in \citet{key-3} (hereafter Z09).
 In this paper, we present data for 
additional $\sim 30$ stars obtained during 2005-2008.
High signal-to-noise ($>$100), high-resolution ($R\sim$55000) 
spectroscopy of all of these stars was carried out with 
the High-Dispersion Spectrograph (HDS: \cite{key-75}) mounted on 
the Subaru telescope. Chemical abundances of Na, Mg, Si, Ca, Ti, Cr, Zn, 
Fe, Ni, Mn, Y and Ba are derived with a homogeneous manner for all 
of these sample. 
The derived abundances are combined with 
abundance data published in literature (SB02; \cite{key-72}) 
resulting in a large sample consisting of $>$200 stars for which 
both high-resolution abundance data and three-dimensional 
velocity components are available. 
Section \ref{sec:sample} describes our sample 
considered in the following 
discussions, which includes stars observed during 2003-2008 or 
published in literature. 
Section \ref{sec:obs} describes our high resolution 
spectroscopy  for a subset of 28 stars obtained in July 2008. 
Data reduction and abundance analysis for all of our 
57 HDS sample are then presented.  
Based on the compiled data-set, Section \ref{sec:results} shows 
abundance trends according to their orbital characteristics. 
Section \ref{sec:disc} discusses implication of these results for the 
formation of the Galactic halo. Finally, Section \ref{sec:conc} 
presents concluding remarks in this work.

\section{The Sample}
\label{sec:sample}
The sample used in our work consists of $\sim 200$ 
metal-poor dwarf and giant stars that are 
presumably belonging to the thin/thick disk or the stellar halo 
of the MW. This sample can be grouped into three subsets of data
based the status of observations and analysis as follows; 
1) objects observed during 2005-2008 with Subaru/HDS, whose 
detailed chemical abundances are derived and calibrated 
in this work ({\it the main sample}), 
2) objects observed in February 2003 with the Subaru/HDS, whose 
abundance analysis was performed in Z09. 
These objects are reanalyzed with the same technique 
as for the main sample for the purpose of keeping 
homogeneity in the analysis method, 
3) objects observed and analyzed in literature. 
Properties of each subset are described in the following subsections. 

\subsection{The main sample}
\label{sec:main}

The major purpose of a set of our observations carried out 
with Subaru/HDS is to obtain accurate chemical 
abundances for presumed outer halo stars that are 
observed and analyzed in a homogeneous manner. 
For this purpose, bright ($V<13$) metal-poor ([Fe/H]$<-1$) stars 
 with $Z_{\rm max}>5$ kpc are selected for observation 
from the catalog of non-kinematically 
selected samples provided by \citet{key-5} and the 
catalog of high-proper motion stars by \citet{key-6} 
and \citet{key-37}, 
all of which include available radial velocities, 
proper motions and distance estimates. All of these 29 stars 
were observed with Subaru/HDS in June and July 2008. 
Details of the observation, analysis of stellar 
atmospheric parameters (effective temperature: $T_{\rm eff}$, 
surface gravity: $\log g$, micro-turbulent velocity: $\xi$) 
and the abundance analysis 
are described in Section \ref{sec:obs}. 

Other two stars were observed in May 2005 (G 64-12 and G 64-37).  
These observations were 
carried out with exceptionally high signal-to-noise ratios 
for the purpose of a Li abundance estimation. 
Since they also have $Z_{\rm max}>5$ kpc and were observed 
with the similar instrumental set-up to the 29 stars described 
above, they are included in the main sample as well.

\subsection{The reanalyzed Subaru/HDS sample}
\label{sec:reanal}

High-resolution spectra for 26 metal-poor dwarfs and giants
 selected by the same criteria as the main sample were 
obtained in February 2003 with Subaru/HDS. These stars 
have been independently reduced and 
analyzed in Z09. 
The observation was carried out with the 
similar spectral coverage, spectral resolution and signal-to-noise 
ratios to the main sample. Details of observations, 
reduction and analysis and results are given in Z09. 
In order to keep homogeneity in the current analysis 
technique, we reanalyzed this subset of data with the same method 
as for the main sample ( Section \ref{subs-1}-\ref{subs-3}). 
The derived values of stellar atmospheric parameters and 
chemical abundance results reasonably agree with each 
other in the two analysis methods as partially reported 
by Z09.  

\subsection{The sample from literature}
\label{sec:lit}

The third subset for which both high-resolution 
abundance analysis and kinematic data are available 
are taken from SB02 and \citet{key-72} (hereafter G03). 
From SB02, data of 46 dwarf stars, 
which do not have a duplicated measurement with 
this work, are employed for the following discussion. 
The abundances of Na, Mg, Si, Ca, Ti, Cr, Fe, Ni, Y and Ba 
were derived in their work from high-resolution ($R\sim$35000-48000) 
and high signal-noise ratio spectra primarily taken 
with the Keck I High Resolution Echelle Spectrometer (HIRES). 
Most of their sample are kinematically peculiar in terms 
of $R_{\rm apo}$, $Z_{\rm max}$ and/or $V_{\phi}$ compared 
to typical inner halo stars. Abundance analysis was 
performed in SB02 with a similar manner to the present work described in 
Section \ref{sec:obs}. 

The abundance data of other 110 stars are taken from 
G03. Their sample consists of metal-poor ([Fe/H]$<0$) disk and 
halo stars with available accurate parallax measurements. 
EWs based on a high-resolution ($R>$50000) and a high 
quality (S/N$>$100) spectra were either measured from the 
data taken with the UVES at Kueyen or assembled from 
literature. In their work, $T_{\rm eff}$ and log$g$ were estimated 
from photometric data while $\xi$ was estimated 
using spectroscopic data similar to this work 
(Section \ref{sec:obs}). Chemical abundance of O, Na, Mg, Si, 
Ca, Ti, Sc, V, Cr, Mn, Fe, Ni and Zn using  
the assembled EWs data were derived adopting these 
atmospheric parameters. An examination for any systematic 
differences in derived abundances in these works with 
those of the present work is performed in Section \ref{sec:crosscal}.

\subsection{Binary stars in the sample}
\label{sec:binary}
In our sample, BD$+$\timeform{04D} 2466 is a known binary from its 
radial-velocity variation \citep{key-78}. As noted in Z09, 
this star shows clear enhancements in [C/Fe], [Y/Fe] and [Ba/Fe].
Thus, the stellar surface abundance of this star 
is most likely affected by 
a mass transfer from an AGB companion. 
 
BD$+$\timeform{01D} 3070 is a suspected binary whose radial velocity 
and an orbital solution were obtained in \citet{key-29}. G 17-25
and G 59-27 are spectroscopic binaries having 
estimates of a robust orbital solution in \citet{key-30}. 
Unlike BD$+$\timeform{04D} 2466, these three stars do not show anomalous 
[Y/Fe] and [Ba/Fe] ratios but show typical values of these ratios 
within a 1$\sigma$ range for each of their metallicity. 
 Therefore, we assume that the surface abundance of 
these three stars is not affected by mass transfers
 from their binary companions.

The amplitudes of radial velocity 
variations of these binary stars are less than 20 km s$^{-1}$. 
Such small variations 
have minimal effects on the calculation of 
kinematics in the Galaxy 
within errors described in Section \ref{sec:kin}. Therefore,
 we include these stars in the following discussion 
assuming that their binary nature does not change 
the inner/outer halo classifications defined below and thus does not 
severely affect results on the overall nature of the inner and 
the outer halos.

\subsection{Kinematics of the sample}
\label{sec:kin}

Three-dimensional velocity components $U$, $V$ and $W$   
as well as orbital parameters including a rotational velocity 
with respect to the Galactic rest frame ($V_{\phi}$), perigalactic and apogalactic 
distances ($R_{\rm peri}$ and $R_{\rm apo}$), 
orbital eccentricity ($e$), and $Z_{\rm max}$ were calculated 
in \citet{key-4} for the Beers et al. sample and we use the 
same method to calculate these kinematical quantities for the 
\citet{key-6} and \citet{key-37} sample. The resulting values 
are tabulated in Table \ref{kin_table}.
The calculations of the orbital parameters as detailed
in \citet{key-4} were performed adopting the St{\"a}ckel-type model 
for the Galactic potential to analytically estimate these values. We 
use the orbital parameters calculated in this manner for the sample 
taken from literature as well, instead of the published values, 
in order to minimize systematic effects in classifying them 
based on kinematics in the following discussion.  

The errors in 
these parameters are calculated by generating a thousand of 
pseudo data of positions, distances and 3-dimensional velocity 
components whose distributions 
are Gaussian with dispersions equal to the observational errors of 
each measured quantity. 
For each of the pseudo data, orbital parameters are
calculated as described above. The Gaussian 
dispersions in the resulting orbital parameters are taken 
as 1 $\sigma$ errors in orbital parameters in the following 
discussions. Median values of errors in $e$, $R_{\rm apo}$, 
$R_{\rm peri}$, and $Z_{\rm max}$ are 0.048, 1.8 kpc, 0.4 kpc 
and 1.8 kpc, respectively.
  
\begin{figure}
  \begin{center}
    \FigureFile(90mm,90mm){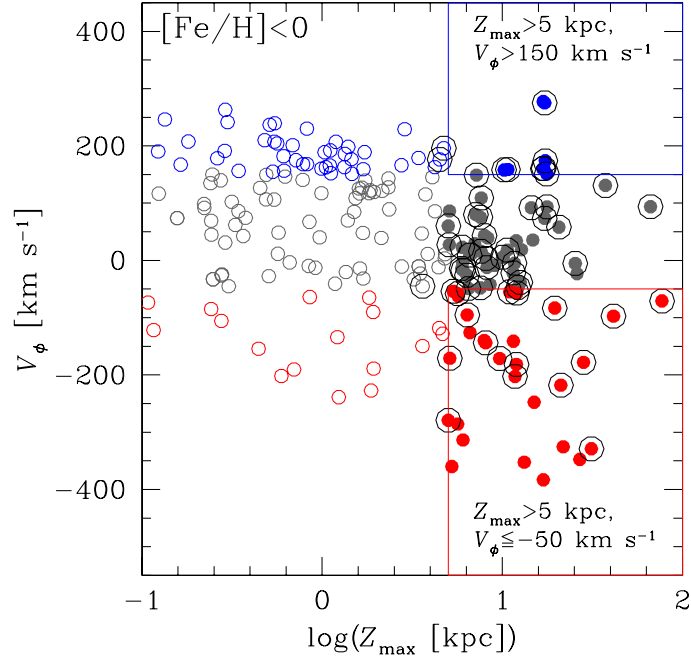}
    \end{center}
  \caption{A $Z_{\rm max}-$$V_{\phi}$ plot for the sample stars with [Fe/H]$<$0. All of the sample stars in this plot have known kinematics (3-dimensional velocity components, distances) and the measured abundances from our high resolution spectroscopy with Subaru HDS or from literature (SB02; G03). {\it Open} and {\it filled} symbols represent stars with $Z_{\rm max}\leqq5$ kpc and $Z_{\rm max}>5$ kpc, respectively. {\it Blue}, {\it gray} and {\it red} colors represent stars with $V_{\phi}>150$ km s$^{-1}$, $-50<$$V_{\phi}\leqq150$ km s$^{-1}$ and $V_{\phi}\leqq-50$ km s$^{-1}$, respectively. The stars observed with Subaru/HDS are shown by symbols encircled with large black open circles.}
\label{vphizmax}
\end{figure}

Figure \ref{vphizmax} is the plot for the combined sample 
on a log($Z_{\rm max}$)-$V_{\phi}$ plane. 
 The sample stars are tentatively divided 
into three different domains in $V_{\phi}$ and two in
$Z_{\rm max}$ in order to investigate any systematic 
differences in chemical abundances of the sample stars 
on their kinematic properties. 
For the $V_{\phi}$ domains, 
$-50<$$V_{\phi}\leqq$150 km s$^{-1}$ ({\it gray} symbols), 
is assumed to represent the range of $V_{\phi}$ typical of 
inner halo stars, while 
$V_{\phi}>$150 km s$^{-1}$ 
({\it blue} symbols) and $V_{\phi}\leqq-50$ km s$^{-1}$ ({\it red} symbols)  
include that of prograde and retrograde outliers, respectively.
This classification is based on recent determinations of 
a modest net prograde rotation (0-50 km s$^{-1}$ with dispersion 
$\sim 100$ km s$^{-1}$) for the inner halo stars \citep{key-4, key-35}.
For the $Z_{\rm max}$ domains, $Z_{\rm max}\leqq5$ kpc ({\it open} 
symbols) and $Z_{\rm max}>5$ kpc ({\it filled} symbols) 
are assumed to be typical $Z_{\rm max}$ domains for the
 ``inner halo'' and the ``outer halo'' candidates.  
The choice of the boundary of $Z_{\rm max}=5$ kpc is 
motivated by the result of \citet{key-35} who 
showed that the average [Fe/H] for their sample stars 
with $Z_{\rm max}>5$ kpc gradually shifts toward lower metallicity 
as stars with higher $Z_{\rm max}$ are selected. 
We note that this $Z_{\rm max}$ criterion results in an 
inclusion of stars with kinematics typical of thick disk stars 
($Z_{\rm max}<1$ kpc and $V_{\phi}>100$ km s$^{-1}$) 
in the inner $Z_{\rm max}$ domain.
Consequently, the sample stars are divided into 
the six kinematic subgroups, whose member stars are represented by 
different symbols in Figure \ref{vphizmax}. 

\begin{figure}
  \begin{center}
    \FigureFile(155mm,155mm){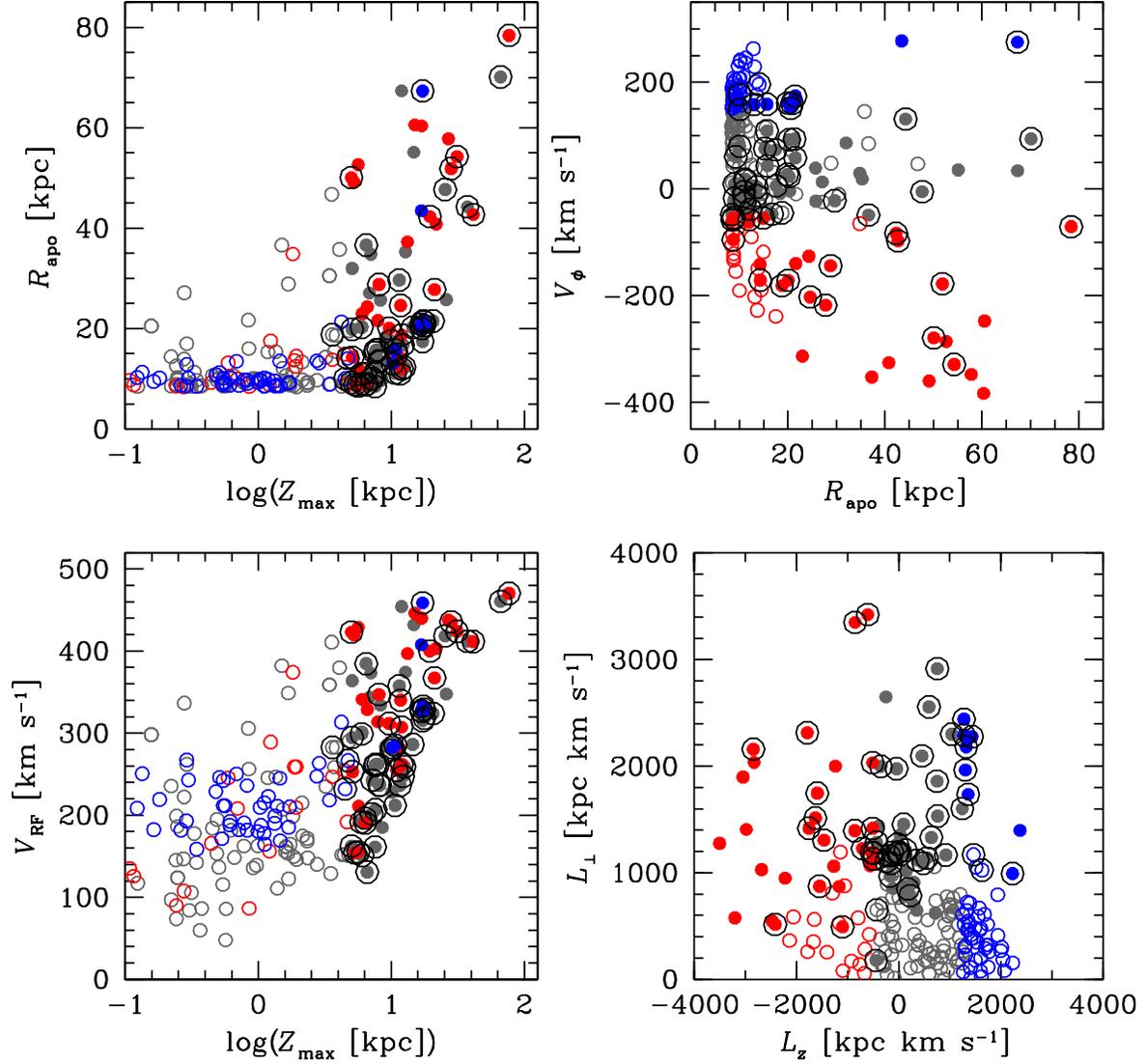}
    \end{center}
\caption{The {\it top-left}, {\it top-right} and {\it bottom-left} panels show correlations between orbital parameters $Z_{\rm max}$-$R_{\rm apo}$, $V_{\phi}$-$R_{\rm apo}$ and $Z_{\rm max}$-$V_{\rm RF}$, respectively, for the sample stars. The {\it bottom-right} panel shows locations of the sample stars on an angular momentum space ($L_{z}$-$L_{\perp}$). The symbols are the same as in Figure \ref{vphizmax}.}
\label{kin_cor}
\end{figure}

Correlations between the orbital parameters for the sample stars 
are shown in Figure \ref{kin_cor}. 
The symbols are the same as the previous figure. These plots 
schematically illustrate whether our conventional classification of the six 
kinematic subgroups using $Z_{\rm max}$ and $V_{\phi}$ fit into 
kinematics of more realistic Galactic components. 
As shown in a $\log Z_{\rm max}$-$R_{\rm apo}$ plane ({\it top-left} panel), 
the criterion of $Z_{\rm max}=5$ kpc as an inner/outer halo boundary results in an 
inclusion of stars with a wide range of $R_{\rm apo}$ ($\sim 8-80$ kpc) 
in the outer $Z_{\rm max}$ domain. It is also noticeable that 
the majority of the sample stars with $Z_{\rm max}\leqq5$ kpc and $V_{\phi}>150$ km s$^{-1}$ 
({\it blue open} symbols) have $R_{\rm apo}\sim 10$ kpc,
which suggests that this kinematic subgroup largely includes 
objects belonging to the thin or thick disk component. 
The {\it top-right} panel shows a plot of the sample stars 
in a $R_{\rm apo}$-$V_{\phi}$ plane. It can be seen that 
as the $R_{\rm apo}$ increases, the sample stars tend to have negative 
$V_{\phi}$. 
A plot of a Galactic rest frame velocity, $V_{\rm RF}$, against
the $\log Z_{\rm max}$ for the sample stars is 
shown in the {\it bottom-left} panel. 
The plot shows that the sample stars with $Z_{\rm max}>5$ 
kpc currently have a wide range of $V_{\rm RF}$ values 
at the solar neighborhood. 
The {\it bottom-right} panel shows locations of the sample stars on an 
angular-momentum space ($L_{z}$-$L_{\perp}$), where 
$L_{z}=RV_{\phi}$ and $L_{\perp}=(L_{x}^{2}+L_{y}^{2})^{1/2}$. 
For the present 
sample that currently resides close to the Galactic plane, $L_{\perp}$ 
is mostly determined by $U$ and $W$ velocity components 
which are coupled with $Z_{\rm max}$. As expected, the sample 
stars with $Z_{\rm max}>5$ kpc predominantly reside in the 
region of $L_{\perp}>1000$ kpc km s$^{-1}$.

\section{Observation and Data Analysis}
\label{sec:obs}

\subsection{Subaru HDS observation}
\label{subs-1}

High-resolution spectroscopy for most of 
the main sample was carried out on 2008 July 27 
and 28 with Subaru/HDS.
Wavelength range of  
4030-6740{\AA} was covered with a gap between two CCDs 
at $\sim5300-5480${\AA}. CCD on-chip binning with 2$\times$2 
pixels was applied for the spectra. Slit of \timeform{0''.70} 
widths is used, which
yields a spectral resolution, measured 
as a FWHM of an emission line profile of a Th-Ar lamp spectrum, 
is $R\sim$55000 at $\sim 5000$ {\AA}. 
Additional three stars (BD$+$\timeform{13D} 2995, G 14-39 and G 20-15) 
were observed in 2008 June in a service program of Subaru/HDS. 
Positions, $V$-band magnitudes, total exposure time, 
signal-to-noise ratio and radial velocity estimates for the sample stars 
are summarized in Table \ref{obslog}.

\subsection{Data reduction}
\label{subs-2}

\begin{figure}
  \begin{center}
    \FigureFile(80mm,80mm){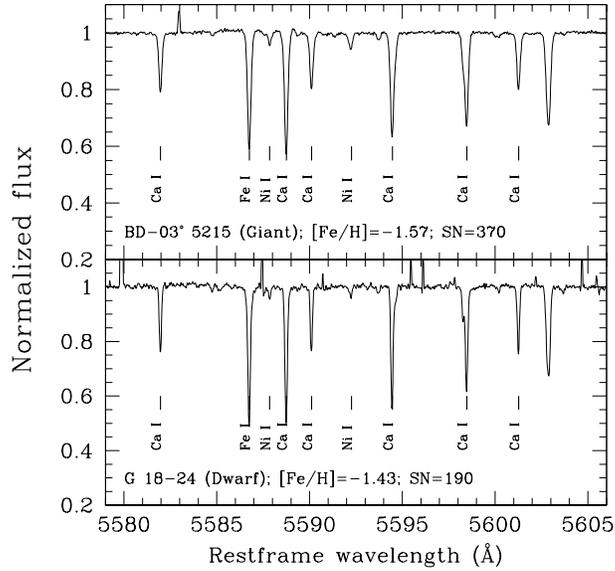}
    \end{center}
\caption{Example of spectra obtained in the observation with Subaru/HDS. 
Top panel shows a spectrum of BD$-$03$^{\circ}$ 5215 (giant) with [Fe/H]=$-1.60$ and S/N$=369$. 
Bottom panel shows a spectrum of G 18-24 (dwarf) with [Fe/H]=$-1.62$ and S/N$=186$.}
\label{exspec}
\end{figure}

The raw data were reduced with standard IRAF routines.
Subtraction of bias frames, flat fielding, cosmic ray rejection, 
background subtraction and wavelength calibration were performed 
for individual frames. When more than one frame are obtained, they are 
combined to yield a single spectrum for each star. Examples of 
reduced spectra, after normalized with the fitted continuum flux, are 
shown in Figure \ref{exspec}. 

\subsection{Equivalent width measurement}
\label{subs-3}

Equivalent widths are measured for absorption lines in the reduced spectra 
by fitting a Gaussian to each feature. The absorption lines 
to be measured and their atomic data (the central wavelength, lower 
excitation potential and $\log gf$) are mainly 
adopted from SB02. Also, we supplement the atomic data for  
the Mg lines, Zn lines and the lines with wavelength $\lesssim4500$ {\AA} 
from the Vienna Atomic Line Database \citep{key-85}. Additionally, we adopt 
one Y II line and two Mn lines from \citet{key-84}. 

For Na lines that are weak and/or 
located close to the neighboring line, a direct integration was used 
for the EWs measurements instead of a Gaussian fitting according to 
the following formula:
\begin{equation}
EW=\int_{\lambda_{\rm min}}^{\lambda_{\rm max}}\frac{F_{\rm cont}-F_{\rm line}}{F_{\rm cont}}d\lambda
\end{equation}
where $\lambda_{\rm max}$ and $\lambda_{\rm min}$ are set to be 
at the wavelength of $\pm 3\sigma$ of the typical Gaussian line profile 
in the present analysis, $F_{\rm line}$ and $F_{\rm cont}$ are a flux 
of the spectrum and an interpolated continuum flux at the line center, 
respectively. Measured EWs, the atomic data with their sources are tabulated in 
Table \ref{ews} (electronic version).

The errors in the equivalent widths ($\sigma_{EW}$) 
measurements are approximated 
with the following formula \citep{key-31}:
\begin{equation}
\sigma_{EW}\sim 1.5 \frac{\sqrt{\mathstrut {\rm FWHM}(\Delta x)}}{S/N}
\end{equation}
where $\Delta x$ is a spectral dispersion in units of {\AA} per pixel. 
For our sample stars, a spectral dispersion is set to
 $\Delta x=0.030${\AA} at 5000 {\AA} in the case of the $2\times 2$ binning was applied. 
As a result, provided that an FWHM in our observation is 
typically $\sim$0.18 {\AA} and S/N=150-500, $\sigma_{EW}$ 
ranges from 0.1 to 0.7 m{\AA}.   

\begin{figure}
\begin{center}
\FigureFile(160mm,80mm){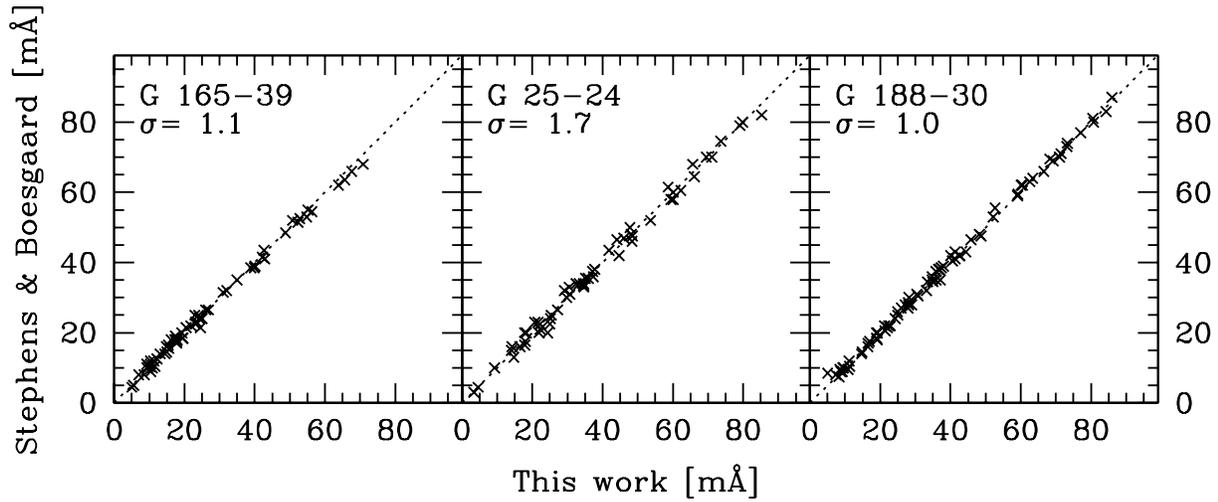}
\end{center}
\caption{Comparisons of measured EWs (m{\AA}) in this work and in SB02 for 
three stars in common, G 165-39 ({\it left}), G 25-24 ({\it middle}) and 
G 188-30 ({\it right}).}
\label{eqcomp}
\end{figure}

The measured EWs are compared with those measured in SB02 
for three stars in common (G 165$-$39, G 25$-$24 and G 188$-$30). 
For these stars, a root mean square (RMS) differences of EWs measured 
in the two studies is $\sigma_{\rm rms}<2$ m{\AA}. 
Given that the internal errors in both measurements are 
in a range $\sim 1-2$m{\AA}, the agreement is excellent. 
Comparisons of EWs for these three stars are shown in Figure \ref{eqcomp}.

\subsection{Abundance analysis}

Abundance analysis was performed using an LTE abundance analysis code 
described in \citet{key-79}. We adopt the model photosphere 
of Kurucz (1993) with a revised opacity distribution function 
(``NEWODF'', \cite{key-77}). We used Uns\"{o}ld's (1955) treatment 
of van der Waal's broadening enhanced by a factor of 2.2 in $\gamma$ 
as in \citet{key-87}.  
Stellar atmospheric parameters are firstly determined. 
Then these values are applied to obtain abundances of individual elements 
from measured EWs.
  
\subsubsection{Stellar atmospheric parameters}
\label{sec:stellaratms}

Effective temperature ($T_{\rm eff}$), surface gravity ($\log g$), 
micro-turbulent velocity ($\xi$) and metallicity ([Fe/H]) 
are estimated with an iterative process 
so that an initial guess of a set of these parameters is consistent with 
 the resulting parameter estimates. In this analysis, we use Fe lines 
with $\log({\rm EW}/\lambda)>-4.8$ to exclude strong lines that cannot 
be well approximated by a Gaussian profile. Note that we use 
stronger Fe lines in the case that only a few adequate EWs measurements 
are available for a sample star.
The resulting values are shown in Table \ref{stpm}. 

\begin{figure}
\begin{center}
\FigureFile(80mm,80mm){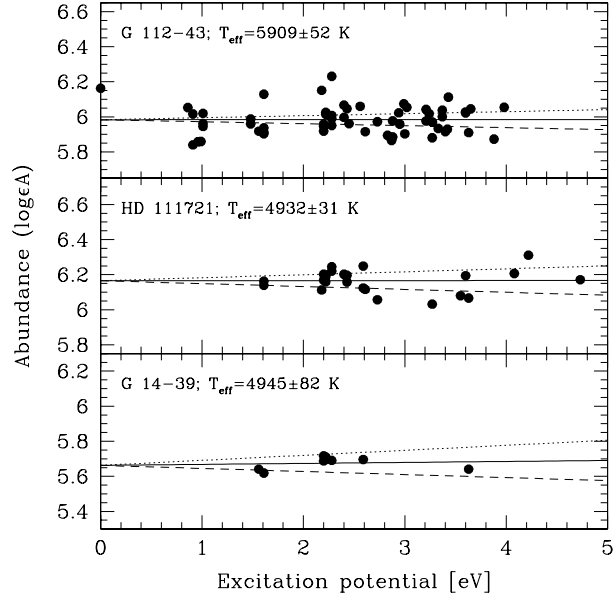}
\caption{Excitation potential ($\chi$) vs log$\epsilon$A. {\it Top}, {\it middle} and {\it bottom} panels show 
the results of G 112-43, HD 111721 and G 14-39, respectively. {\it Dotted} and {\it dashed} lines 
display slopes when $T_{\rm eff}$ is changed by an amount of the error to the positive and negative direction, 
respectively.}
\label{excp_ab}
\end{center}
\end{figure}

$T_{\rm eff}$ was obtained from EWs of $\lesssim 70$ Fe I lines 
so that a trend in derived Fe abundances from each Fe I 
line as a function of their excitation potentials ($\chi$) is 
minimized. 
Three panels of Figure \ref{excp_ab} show
 example plots of a $\log\epsilon A$
abundance\footnote{$\log\epsilon A=\log (N_{A}/N_{H}) +12$} against 
$\chi$ (eV) for three stars in our sample. 
Each panel displays data points and a linear fit ({\it solid} line) to 
these data when the best $T_{\rm eff}$ estimate was applied. 
{\it Dotted} and {\it dashed} lines 
show slopes when $T_{\rm eff}$ is changed by an amount of the error 
to the positive and negative direction, respectively.
Typically, magnitudes of the slopes can be minimized to $<0.01\pm 0.02$ dex eV$^{-1}$, 
where an error in the slope estimate corresponds to $T_{\rm eff}$ of 
20-100 K.

\begin{figure}
\begin{center}
\FigureFile(80mm,80mm){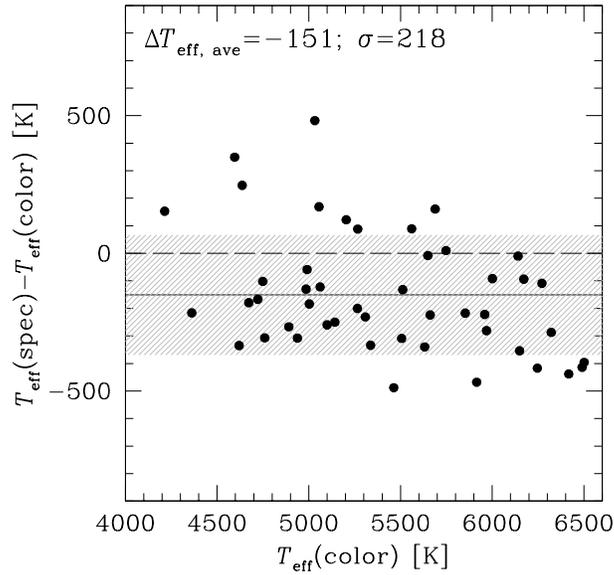}
\caption{Deviation in $T_{\rm eff}$ estimated in our work, $T_{\rm eff}$(spec), from that obtained with ($V-K$) color, $T_{\rm eff}$(color).}
\label{color_temp}
\end{center}
\end{figure}

The values of $T_{\rm eff}$ obtained with this method 
are compared with those derived from a $V-K$ color. 
We use $V$ and $K$ magnitudes from the SIMBAD database and $E(B-V)$ 
values, taken from either the catalog of \citet{key-5} or 
\citet{key-6}, those calculated iteratively in the method described 
in \citet{key-5}. The $E(B-V)$ values were used to correct 
the $V$ magnitude for 
an interstellar extinction adopting a relation,  
$A_{V}=R_{V}E(B-V)$ where $R_{V}=3.1$. The calibration of 
\citet{key-32} and \citet{key-33} for dwarf and giant stars, 
respectively, are used to estimate $T_{\rm eff}$ from  
the extinction corrected $V-K$ colors ($(V-K)_{0}$), where $K$-band 
magnitudes are taken from the 2MASS catalog \citep{key-57}. 
$T_{\rm eff}$ estimated using Fe I lines is lower 
than that estimated from $(V-K)_{0}$ by $\Delta$$T_{\rm eff}
\sim 150$ K on average (Figure \ref{color_temp}). The higher 
$T_{\rm eff}$ estimates from the $(V-K)_{0}$ color than those from 
an $\chi-\log\epsilon A$ plot is previously reported by \citet{key-50}. 
In their sample, negative 
trends in $\chi-\log\epsilon A$ plot are observed when the $T_{\rm eff}$ 
derived from the $(V-K)_{0}$ colors were applied for the abundance 
estimates.

In this work, we adopt the $T_{\rm eff}$ estimation from Fe I lines 
rather than the color temperature because this method can be reasonably 
applied to all of our HDS sample. 
Furthermore, the color temperatures are sensitive to reddening 
values as well as errors in photometry that are typically 
0.01-0.04 mag in $K$-bands. 
Additionally, spectroscopic $T_{\rm eff}$ allows us calibration 
with other spectroscopic work such as SB02. 
In the later section, we show that the spectroscopic $T_{\rm eff}$ 
in this work and in SB02 reasonably agree each other.
Indeed, as shown in Section \ref{sec:error}, 
the effect of change in $T_{\rm eff}$ is comparable to typical errors 
in abundance analysis 
($\Delta$[X/H]$\sim 0.1$ dex) and, in most cases, is canceled out when we 
normalize [X/H] values with the Fe abundances.

$\log g$ was obtained so that 
the abundances derived from Fe I and Fe II are consistent 
within 0.02 dex. This analysis is based on an assumption that 
the derived abundances from neutral and ionized species 
should be the same. 
We note, however, using the estimated $\log g$ values, 
the resulting abundances of Ti and Cr from their neutral 
and ionized species are inconsistent by 0.1-0.4 dex (Section \ref{sec:abu}).

Finally, $\xi$ was obtained so that a trend of derived 
abundances from individual Fe I lines with 
their equivalent widths is minimized. 

\begin{figure}
\begin{center}
\FigureFile(150mm,150mm){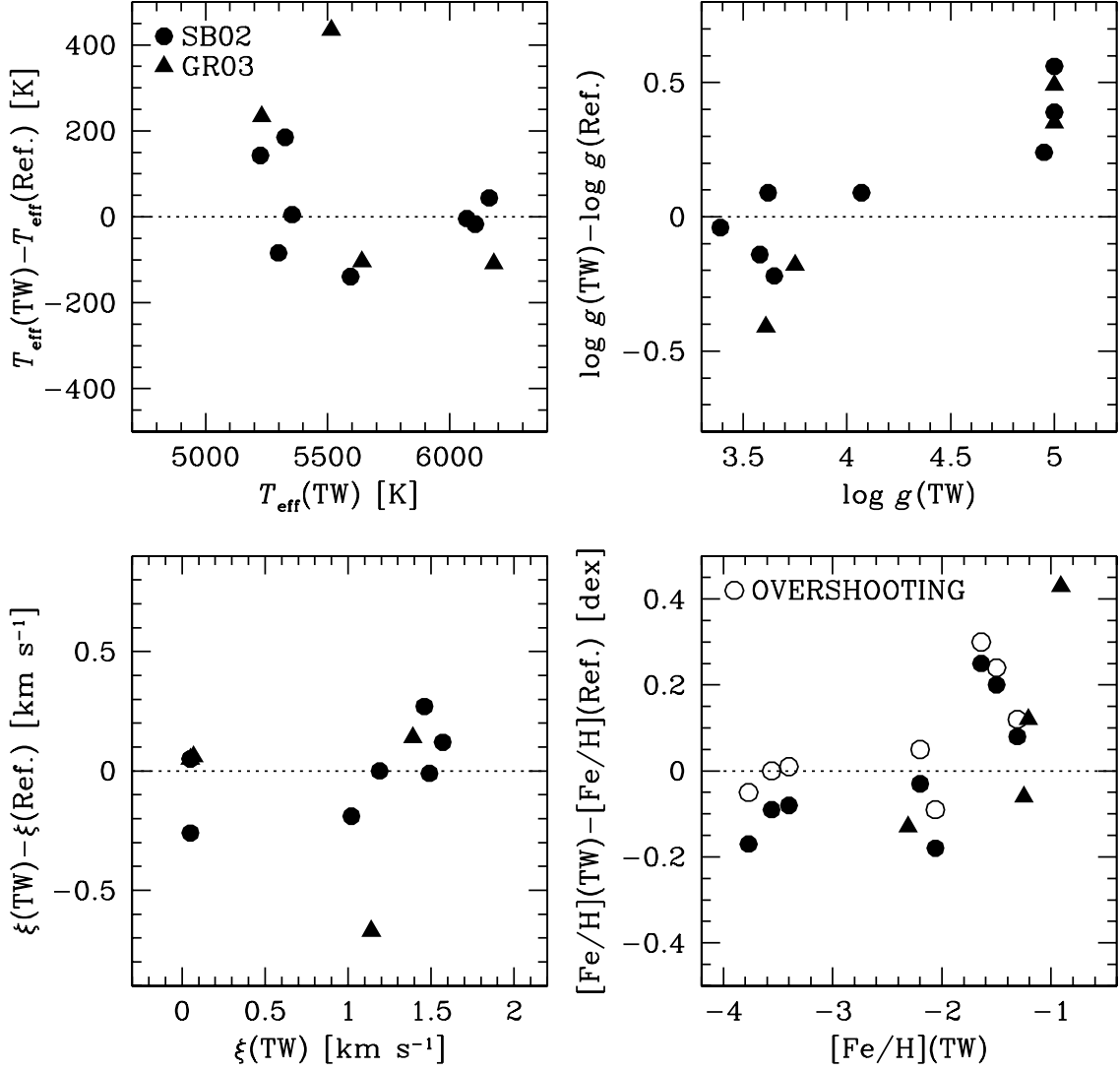}
\end{center}
\caption{Deviation in the stellar parameters, $T_{\rm eff}$, $\log g$, 
$\xi$ and [Fe/H], estimated in this work (TW) from those estimated 
in SB02({\it filled circles}) and G03({\it filled triangles}). For 
the comparison of the [Fe/H] estimates ({\it bottom-right} panel), deviations when 
the overshooting model is applied are shown by {\it open circles}. 
(Section \ref{sec:stellaratms})}
\label{comp_sb_stpm}
\end{figure}

The derived stellar parameters are 
compared with those derived in SB02 for eight objects analysed in common (G 15-13, G 165-39, G 166-37, 
G 188-30, G 238-30, G 25-24, G 64-12 and G 64-37) and in G03 for four
objects analysed in common (G 17-25 or HD 149414, G 43-3 or HD 84937,
 HD 111980 and HD 134439).
 We note that G 43-3 is supplemented to our original 
sample from the archival data of Subaru/HDS \footnote{Based on data collected at Subaru 
Telescope and obtained from the SMOKA, which is operated 
by the Astronomy Data Center, National Astronomical Observatory of Japan.
 \citep{key-83}} in order to better examine systematic differences in derived 
parameters between our work and G03. 
The differences in each parameter (this work - literature) for the sample stars are plotted in 
Figure \ref{comp_sb_stpm} and tabulated in Table \ref{comp_stpm_table}. 
In each panel of Figure \ref{comp_sb_stpm} ($T_{\rm eff}$: {\it top-left}, $\log g$: {\it top-right}, 
$\xi$: {\it bottom-left} and [Fe/H]: {\it bottom-right}), {\it circles} and {\it triangles} 
represent comparisons with SB02 and G03, respectively. 
In Table \ref{comp_stpm_table}, mean differences ($\Delta_{\rm ave}$) and RMS scatters 
($\sigma$) around the mean in each parameters are indicated 
for the comparisons with SB02 and G03.

As can be seen, a mean difference in $T_{\rm eff}$ between SB02 and 
this work is small (17 K) and within the errors in 
our temperature determination. Also, the derived $\log g$ values in this work
reasonably agree with SB02. The agreement of $\xi$
with those derived in SB02 is excellent within scatter $<$0.2 km s$^{-1}$. 

On the other hand, the average differences in these parameters 
from G03 are large, mostly due to G 17-25, which is known 
spectroscopic binary \citep{key-30}. 
In particular, the derived $T_{\rm eff}$ value for G 17-25 is  
significantly different from that derived in G03 by $\Delta>400$ K.
 Such discrepancy could be due to the difference of the analysis 
technique that G03 used $B-V$, $b-y$ colors and a H$_{\alpha}$ line 
profile to estimate $T_{\rm eff}$ while this work utilizes Fe I lines similar to 
SB02. Indeed, G03 reported that G 17-25 is an outlier showing 
significantly different $T_{\rm eff}$'s estimated from $B-V$ and $V-K$.
Accordingly, the difference in the $\log g$ values is large ($0.49$ dex) for this object.  
Excluding this object, the agreement with G03 is reasonable as 
shown in the last low of Table \ref{comp_stpm_table} except that 
$\xi$ is lower for HD 111980 in this work by 0.67 dex from the value derived 
in G03.

The comparison of the [Fe/H] estimates with SB02, except for the 
three objects discussed below,shows that 
the [Fe/H] estimate in this work is systematically lower than that in SB02 
up to $\sim 0.19$ dex. This lower [Fe/H] in this work is partly
due to our use of the model stellar atmosphere without assumption of 
convective overshooting while SB02 utilize models which assume the overshooting 
\citep{key-79}.  
To quantify this effect, 
we perform the [Fe/H] estimate using the overshooting model as in SB02. 
The results are plotted by open circles in the bottom-right panel of 
Figure \ref{comp_sb_stpm}. The resulting [Fe/H] values are higher by 
$\sim 0.1$ dex than the original values and the difference from the 
estimate of SB02 now became small for all stars with [Fe/H]$<-2$.  
On the other hand, the 3 objects (G 15-13, G 166-37 and G 188-30) 
show higher [Fe/H] values in this work than in SB02 by up to 0.2 
dex. Since all of these stars show low $\xi$ values, we suspect 
that different modelings for a line broadening in this work 
 and in SB02 are responsible for the [Fe/H] discrepancies. 
The [Fe/H] is in good agreement with G03 except for G 17-25, 
whose adopted other stellar parameters are significantly 
different from those of G03.

\subsubsection{Abundance}
\label{sec:abu}

The abundances of Na, Mg, Si, Ca, Ti, Mn, Fe, Ni, Zn, Y and Ba for each 
star are calculated adopting the stellar parameters estimated above. 
The abundance estimated from individual lines are averaged over a number 
of detected lines. This value of $\log \epsilon A$ is normalized with 
the solar abundances ($\log \epsilon A_{\odot}$) \citep{key-34}, which yields a value of
[X/H] for each species. The derived abundances in [X/H], errors ($\sigma$) 
and number of lines ($N$) used in the 
abundance estimates are tabulated in Table \ref{tab:ab1}-\ref{tab:ab3}.
In Table \ref{tab:ab4} and \ref{tab:ab5}, Fe-normalized abundance ratios relative to 
solar one, [X/Fe], are listed.  

For the element for which both neutral and singly-ionized 
species are detected for a single element, we took an average of abundances 
derived from individual species weighted by the variance of the mean 
for each species to get the [X/Fe] 
values. 

\begin{figure}
\begin{center}
\FigureFile(80mm,80mm){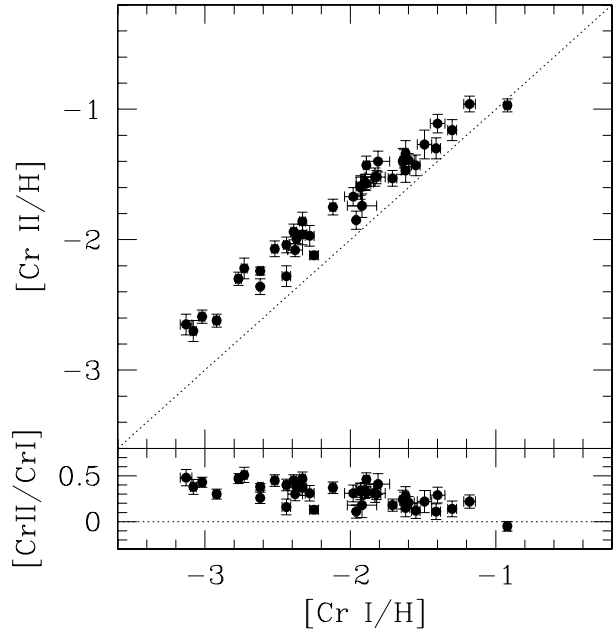}
\caption{{\it Upper panel}: Cr abundances estimated from 
Cr II lines ([Cr II/H]) as a function of those estimated 
from Cr I lines ([Cr I/H]). {\it Lower panel}: Differences in the 
two estimates ([Cr II/H]$-$[Cr I/H]$=$[Cr II/Cr I]) as a 
function of [Cr I/H]}
\label{cr1_cr2}
\end{center}
\end{figure}

We note that a large discrepancy between Cr abundances 
obtained from lines of neutral (Cr I) and singly-ionized (Cr II)
 species is observed for our main sample as illustrated 
in Figure \ref{cr1_cr2}. In this figure,
the upper panel shows Cr abundances obtained from Cr II lines ([Cr II/H]) 
as a function of those obtained from Cr I lines ([Cr I/H]). The 
lower panel shows differences in the two estimates 
([Cr II/Cr I]) as a function of [Cr I/H].
In our main sample, [Cr/Fe] ratios estimated from Cr I 
are lower than those estimated from Cr II by 0.24 and 0.36 dex on 
average in the metallicity range $-2<$[Fe/H]$\leqq-1$ and 
$-3<$[Fe/H]$\leqq-2$, respectively. In most cases, Cr abundances
derived from Cr I are assigned more weight than those from Cr II 
in this study, because a larger number of lines are detected 
yielding a smaller variance of the mean for Cr I. 
Such discrepancy was already reported for
 metal-poor stars with $-4<$[Fe/H]$<-2$ studied by 
\citet{key-50}. \citet{key-51} suggest that the discrepancy is
possibly due to a NLTE effect which corresponds to over ionization 
of Cr I, although an exact cause of this phenomenon is still unclear. 
The discrepancy also presents in our subsample taken from literature 
(SB02 and G03). 

\begin{figure}
\begin{center}
\FigureFile(80mm,80mm){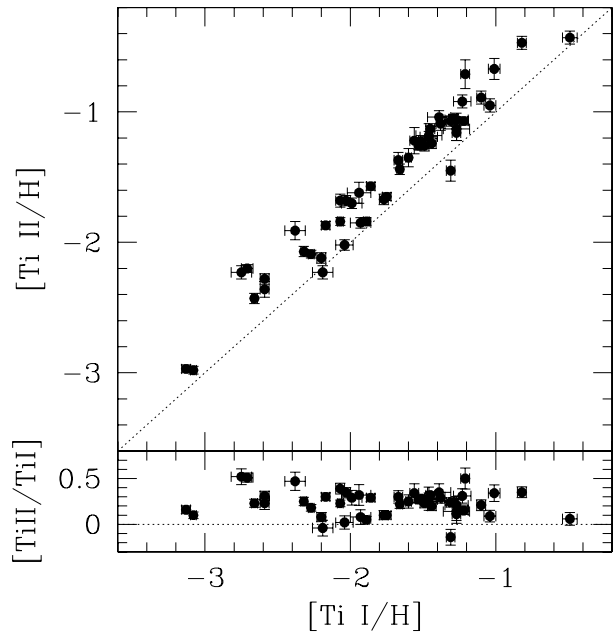}
\caption{Same as Figure \ref{cr1_cr2} but for Ti.}
\label{ti1_ti2}
\end{center}
\end{figure}

Similarly, Ti abundances estimated from lines of neutral (Ti I) 
and ionized (Ti II) species are different by $\sim 0.25$ dex on average, 
as illustrated in Figure \ref{ti1_ti2}. 
Such a discrepancy could be minimized if $\log g$ was adopted 
so that the Ti I and Ti II lines yield similar Ti abundances as 
was performed by SB02. 
Therefore, a caution must be needed when 
comparing the [Cr/Fe] and [Ti/Fe] 
abundance results in this study with those from literature 
(Section \ref{sec:crosscal}).

For Mn and Ba abundances, the hyper-fine structure (HFS)
was taken into account in the abundance calculation.
The data on the HFS of Mn I lines are taken from \citet{key-88}
\footnote{The HFS lists are available at http://kurucz.harvard.edu/LINELISTS/GFHYPER100/}.
 For the HFS of Ba, the line list of \citet{key-11}, assuming the isotopic abundance 
of r-processes estimated for the solar system, was used. 

\subsection{Error estimate} 
\label{sec:error}

For given $T_{\rm eff}$, $\log g$ and $\xi$ values, 
uncertainties in abundances ($\log \epsilon A$)
are taken as RMS line-to-line scatter (standard deviation
: $\sigma_{\log \epsilon A}$) of 
abundances estimated from individual lines as divided by a square-root of 
the number of lines detected for the species. In the case of only 
one or two line are detected for a particular element, 
the uncertainties are taken to be equal to $\sigma_{\log \epsilon A({\rm Fe I})}$ for that 
sample star. 
The values of $\sigma_{\log \epsilon A({\rm Fe I})}$ is typically 0.05-0.1 dex. 
Thus, the error in an average of Fe abundances 
derived from each of $<70$ Fe I lines results in 0.01-0.02 dex. 
Using these abundance errors, errors for abundance ratios [X/H] are 
calculated as a square root of the quadratic sum of the abundance 
errors and the errors in the Solar abundance.
For the Fe-normalized ratio [X/Fe], the error is a square root of 
the quadratic sum of errors in [X/H] and [Fe/H]. The errors
are tabulated along with abundances 
in Table \ref{tab:ab1}-\ref{tab:ab5}.

\begin{figure}
\begin{center}
\FigureFile(160mm,160mm){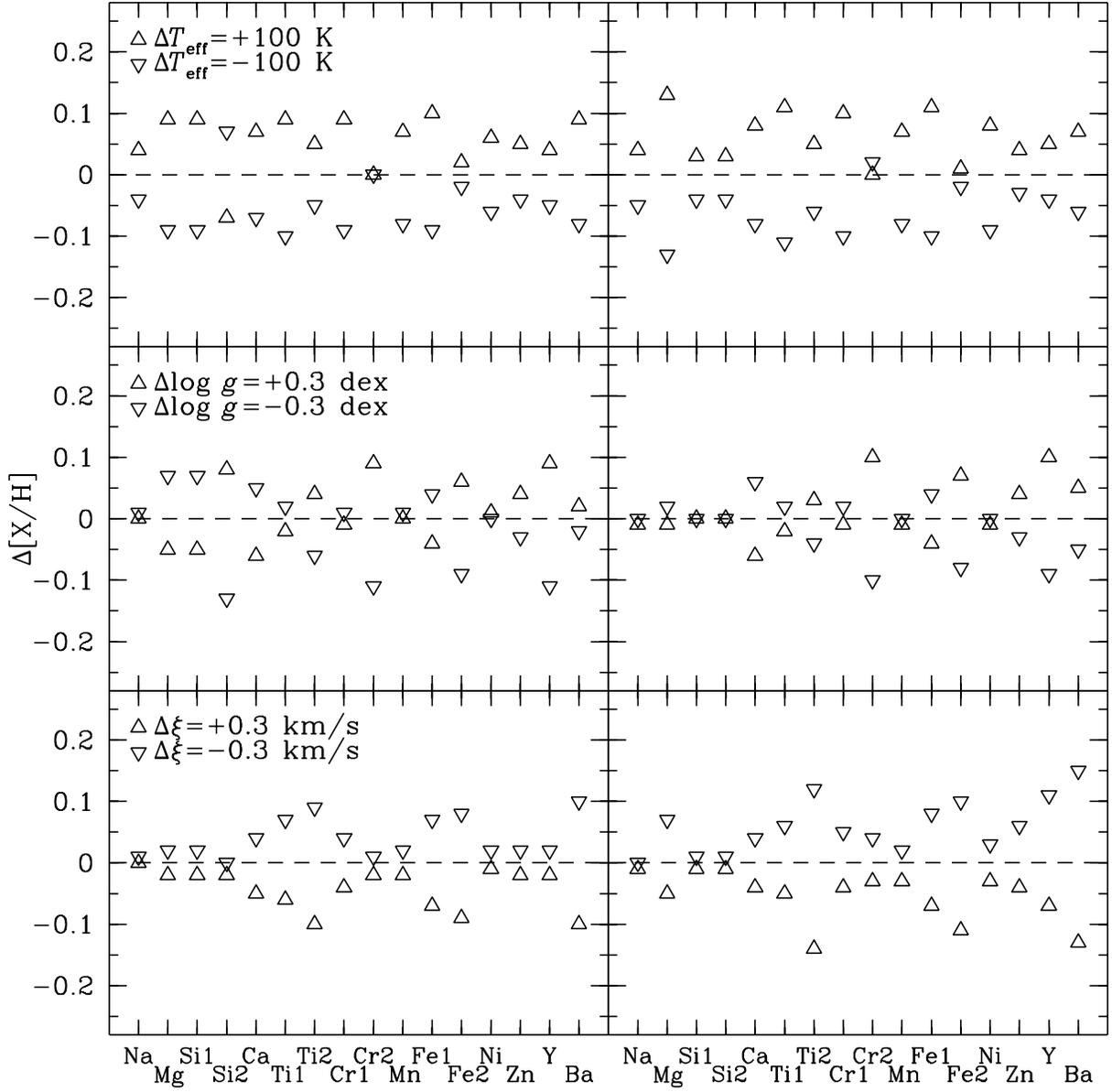}
\end{center}
\caption{Abundance deviation ($\Delta$[X/H]) for 
a dwarf, G 275$-$11 ({\it left}), and a giant star, HD 111980 ({\it right}),
 in our sample when 
stellar atmospheric parameters $T_{\rm eff}$ ({\it top}), $\log g$ ({\it middle})and $\xi$ ({\it bottom})
are changed by $\pm 100$K, $\pm 0.3$ dex and $\pm 0.3$ km s$^{-1}$, 
respectively. {\it Triangles} show the results of changes in parameters 
to positive direction and {\it inverted triangles} show those to the negative direction.}
\label{dev}
\end{figure}

A largest contribution to uncertainties in derived 
[X/H] is expected to be due to errors in adopted 
stellar atmospheric parameters. In order to estimate 
uncertainties accompanied by this effect, 
changes in abundances caused by change in adopted stellar parameters 
are examined for two stars of our main sample. 
Figure \ref{dev} illustrates abundance deviation ($\Delta$[X/H]) for 
a dwarf (G 275-11) and a giant star (HD 111980) in our sample when 
stellar atmospheric parameters $T_{\rm eff}$ ({\it top}), $\log g$ ({\it middle})and $\xi$ ({\it bottom})
are changed by $\pm 100$K, $\pm 0.3$ dex and $\pm 0.3$ km s$^{-1}$, 
respectively. In most cases, $\Delta$[X/H] is less than $\sim0.1$ dex, 
which is comparable to or larger than the errors estimated from 
the line-to-line scatter. 
However, such abundance deviations 
tend to be canceled out when they are expressed as a ratio to the [Fe/H]. 
The largest deviation, $\Delta$[X/H]$\sim 0.15$ dex, is found for  
[Ba/H] when $\xi$ is 
changed by $\pm 0.3$ km s$^{-1}$. From these consideration, 
a maximum error of 0.15 dex for Ba and 0.1 dex for other elements 
should be taken into account in interpreting the abundance results.
Systematic errors in the analysis method are checked by 
comparing the abundances in this work with those obtained 
in Z09 for 26 stars independently analyzed in both works.  
RMS differences in the [X/Fe] ratios for these objects are in the ranges
of 0.07 to 0.14, that are similar to the internal errors discussed 
above. 

\subsection{Cross calibration for the abundance ratios with literature}
\label{sec:crosscal}

\begin{figure}
\begin{center}
\FigureFile(120mm,120mm){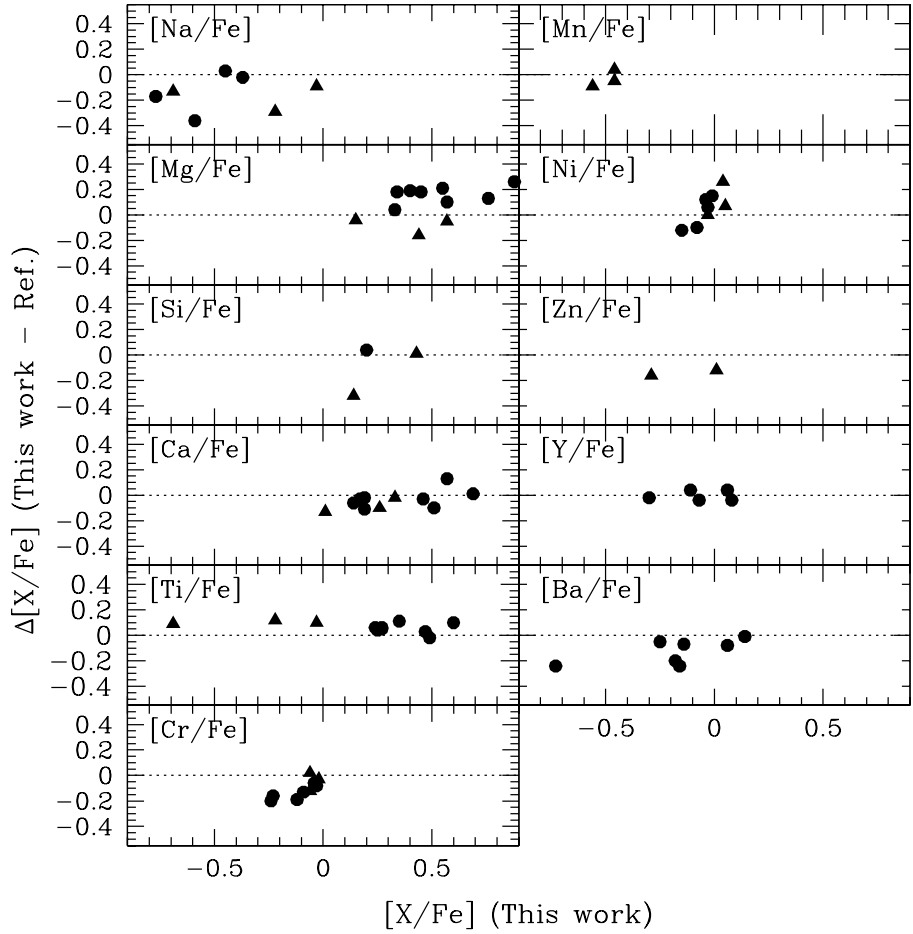}
\end{center}
\caption{Differences in abundance ratios [X/Fe] derived in literature and in this work. {\it Circles} show those derived in SB02 for eight stars (G 15-13, G 165-39, G 166-37, G 188-30, G 238-30, G 25-24, G 64-12, G64-37) and {\it triangles} show those derived in G03 for four stars (G 17-25, G 43-3, HD 111980, HD 134439)}
\label{crosscal}
\end{figure}

The derived abundances for the main sample (Section \ref{sec:main})
 and the reanalyzed sample 
(Section \ref{sec:reanal}) are combined with samples 
from SB02 and G03 as described in Section \ref{sec:lit}, 
which includes smaller $Z_{\rm max}$ stars.
Systematic differences 
arising from differences in the analysis procedures are checked 
using sample stars commonly analyzed in this work and the literature. 
Figure \ref{crosscal} shows the 
differences in derived abundance 
ratios between those in the literature and in this work. 

In this work, 
eight stars are analyzed in common with SB02 ({\it circles} in Figure \ref{crosscal}, 
Mn and Zn have not been measured in SB02) 
as mentioned in Section \ref{sec:stellaratms}.
For most of the elements analyzed, derived abundance ratios in SB02 
are in reasonable agreement with this work within the errors of the measurements, 
except for Na, Cr and Ba.
This work derived lower [Na/Fe], [Cr/Fe] and [Ba/Fe] ratios by $0.16$ dex 
on average than in SB02. The discrepancy in [Na/Fe] ratios in this work and in SB02 
is most likely due to the weakness of the measured lines, whose EWs are 
typically $<10${\AA} and thus, abundance errors of $\sim 0.2$ dex 
is expected in our analysis. 
The discrepancy in [Cr/Fe] is also expected from the large errors in 
abundance ratios caused by taking average of [Cr/H] derived from 
neutral and ionizing species of Cr ([Cr I/H] and [Cr II/H], respectively) 
that are systematically different as described in Section \ref{sec:abu}.  
Such discrepancy is also seen in the sample of SB02 as well. 
The final [Cr/H] is derived by taking weighted average of 
[Cr I/H] and [Cr II/H]. 
In this analysis, difference in the weighting factors, which is determined by 
the number of lines for each species, 
between present work and SB02 may cause the large scatter in final abundances. 
The lower [Ba/Fe] could be due to the hyper fine structure of Ba 
lines considered in this work. 
Except for these three elements, the mean difference for each element is
 less than 0.12 dex. 

For the four stars (G 17-25, G 43-3, HD 111980 and HD 134439) 
analysed in common with G03 ({\it triangles} in 
Figure \ref{crosscal}, Y and Ba have not been measured in G03), 
large discrepancies 
in [Na/Fe] and [Si/Fe] are found for G 17-25 (spectroscopic binary), 
for which we reported in Section \ref{sec:stellaratms} 
that the large discrepancies in 
derived stellar parameters are seen; G03 adopted 
$T_{\rm eff}=5080$ K estimated from a $B-V$ color and [Fe/H]$=-1.34$ while this work 
adopted a spectroscopic $T_{\rm eff}$ of 5515 K and [Fe/H]$=-0.91$ 
(Table \ref{comp_stpm_table}). 
Apart from this object, an
agreement is good on average within 0.12 dex except for Zn. We note 
that this systematic offset in Zn could cause an 
artificial difference in abundance 
between the inner and the outer halo samples 
since most of the outer 
halo stars came from our observation while the inner halo sample 
mostly came from G03. In particular, small systematic 
offset (0.09 dex) in [Mg/Fe] ratio, which is one of the most important 
key $\alpha$ elements for the comparison between different halo populations, 
should be considered in interpreting the final result. We note that another 
$\alpha$ element abundance, [Ca/Fe], shows a good agreement
between the two works.

\section{Results}
\label{sec:results}

In the following subsections, we show the resulting abundance ratios 
([X/Fe]) for all sample stars described in Section \ref{sec:sample}. 
For the sample stars from literature, the [X/Fe] values are 
taken from published values, assuming that  
any systematic differences in derived [X/Fe] ratios by different methods 
are sufficiently small compared to the typical errors in the [X/Fe] 
in this work ($\sim$0.1 dex). We caution that abundances of some elements, 
especially for Na, Cr, Ba, etc. strongly depend on various assumptions 
employed in different analysis methods, such as treatment of NLTE effects. 
In the following, we present results from our LTE 
abundance analysis and do not apply the correction for the NLTE effects.

For clarity, we refer the subsets of the sample with $Z_{\rm max}\leqq 5$ kpc 
and $Z_{\rm max}>5$ kpc as ``{\it inner halo sample}'' 
and ``{\it outer halo sample}'', 
respectively. Since the proposed inner and outer halo components
broadly overlap each other \citep{key-35}, the boundary of $Z_{\rm max}=5$ kpc 
is set as a conventional criterion and, therefore, does not
necessary reflect the realistic natures of these components. 
We discuss the validity of this conventional criterion
in Section \ref{sec:boundary}.

\subsection{[Fe/H]-[X/Fe] relations}

\begin{figure}
  \begin{center}
    \FigureFile(160mm,160mm){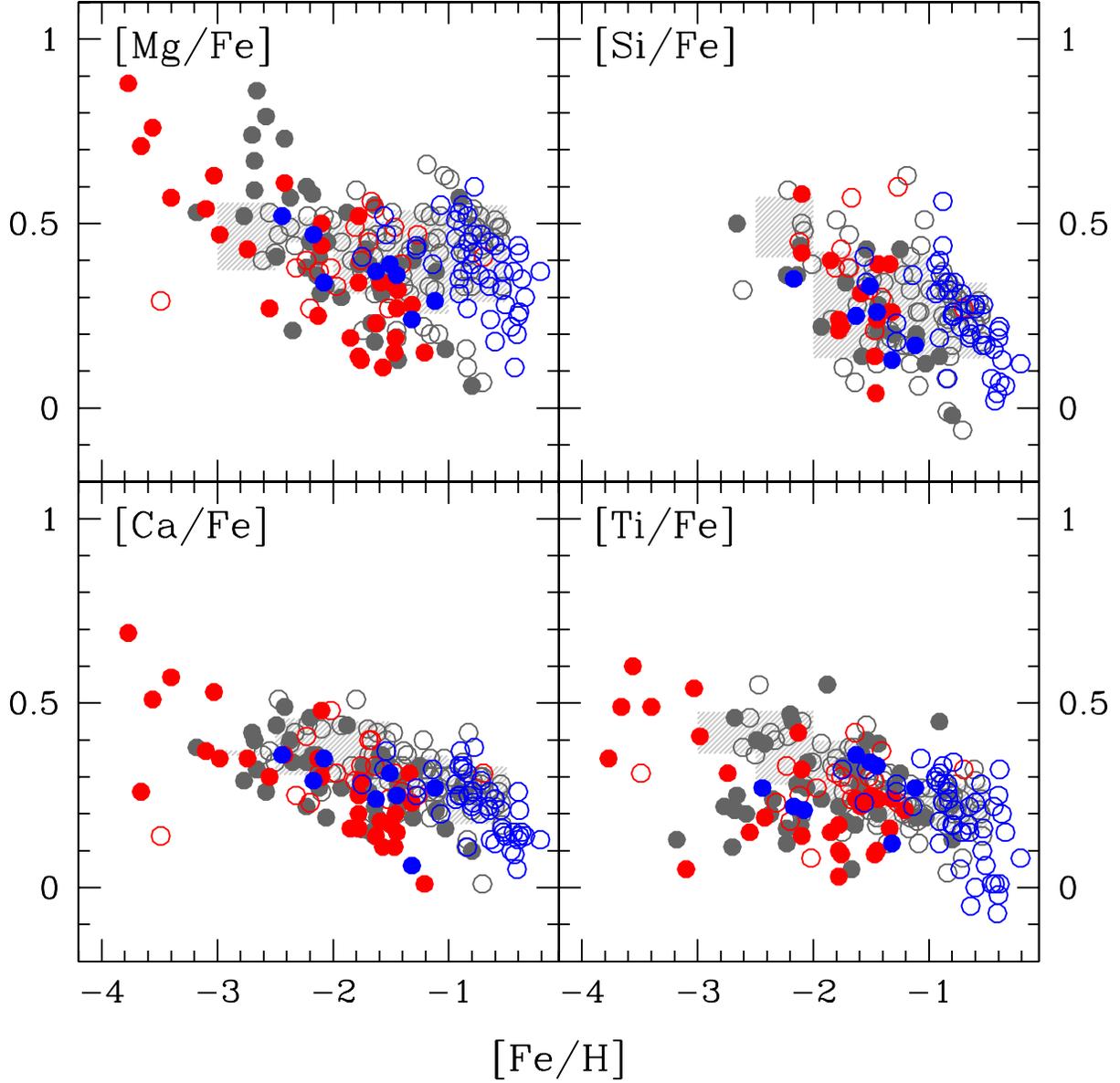}
  \end{center}
  \caption{Mg, Si, Ca and Ti abundance ratios as a function of 
[Fe/H]. As in Figure \ref{vphizmax}, {\it open} and {\it filled} symbols represent stars with $Z_{\rm max}\leqq5$ kpc and $Z_{\rm max}>5$ kpc, respectively. {\it Blue}, {\it gray} and {\it red} colors represent stars with $V_{\phi}>150$ km s$^{-1}$, $-50<$$V_{\phi}\leqq150$ km s$^{-1}$ and $V_{\phi}\leqq-50$ km s$^{-1}$, respectively. The shaded region highlights average and a 1$\sigma$ scatter of the abundance ratios for the sample stars with a
typical inner halo kinematics, which is assumed to be
$Z_{\rm max}\leqq5$ kpc and $-50<$$V_{\phi}\leqq150$ km s$^{-1}$.}
\label{feh_afe}
\end{figure}

\begin{figure}
  \begin{center}
    \FigureFile(80mm,80mm){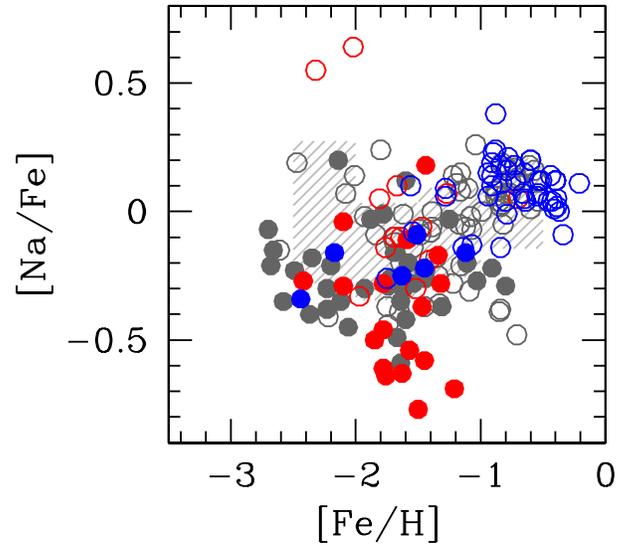}
  \end{center}
  \caption{[Na/Fe] as a function of [Fe/H]. The symbols 
are the same as in Figure \ref{feh_afe}.}\label{feh_nafe}
\end{figure}

\begin{figure}
  \begin{center}
    \FigureFile(160mm,160mm){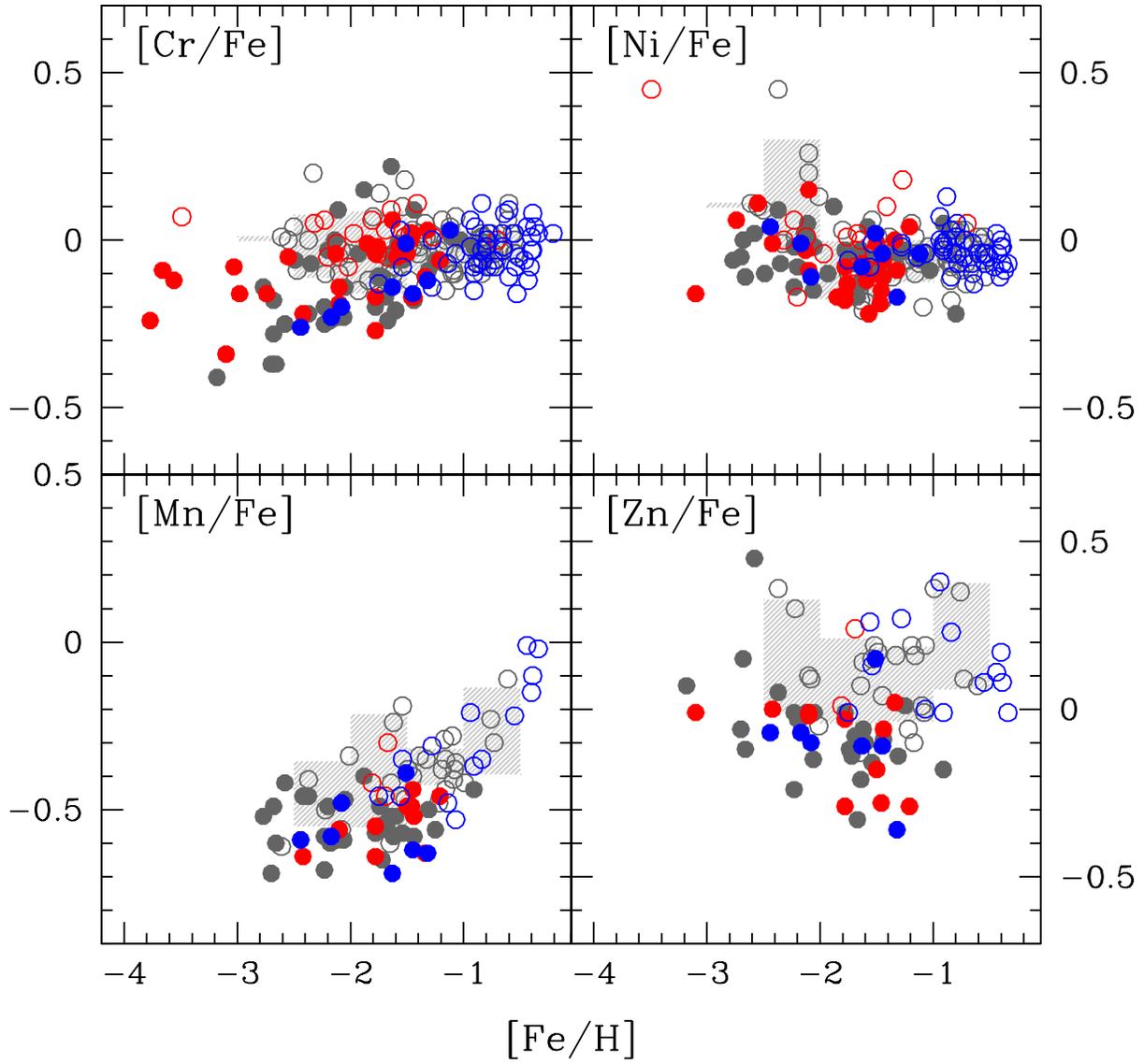}
  \end{center}
  \caption{Cr, Ni, Mn and Zn abundance ratios as a function of 
[Fe/H]. The symbols are the same as in Figure \ref{feh_afe}.}
\label{feh_fepfe}
\end{figure}

\begin{figure}
  \begin{center}
    \FigureFile(160mm,160mm){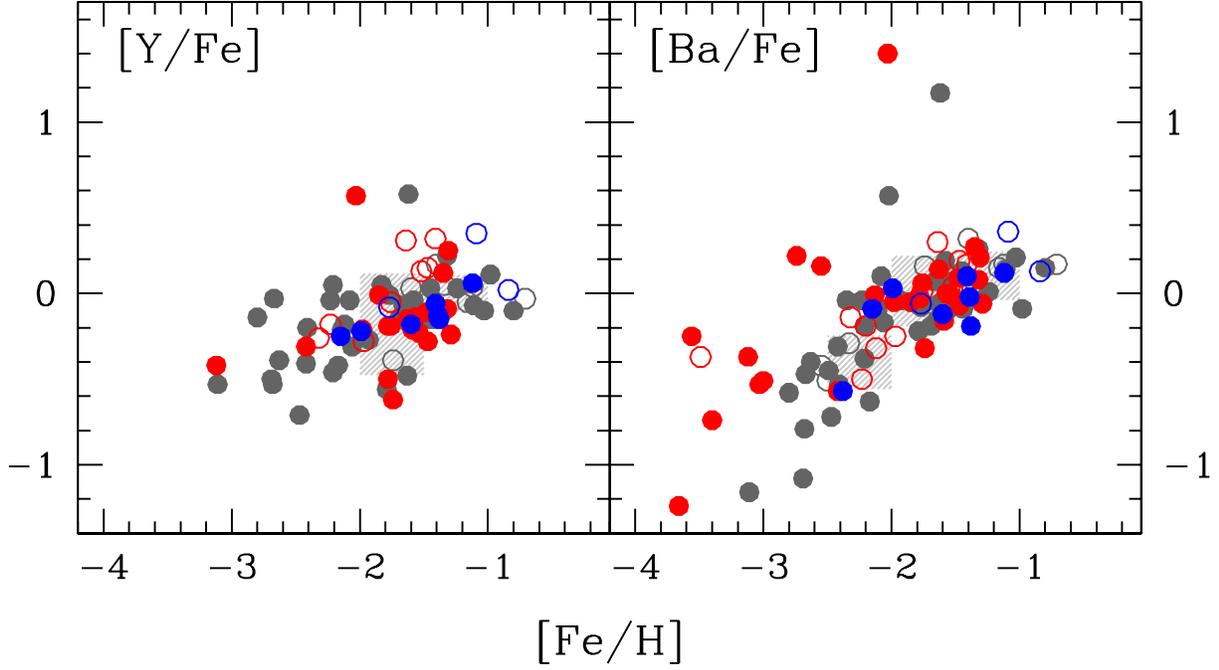}
  \end{center}
  \caption{Y and Ba abundance ratios as a function of 
[Fe/H]. The symbols are the same as in Figure \ref{feh_afe}. We note that 
one of the two objects having exceptionally high [Y/Fe] and [Ba/Fe] is 
a binary star (BD$+$\timeform{04D} 2466, [Fe/H]$=-2.16$) while binary 
nature of another object (G 18-24, [Fe/H]$=-1.62$) is unknown \citep{key-30}.} 
\label{feh_ncfe}
\end{figure}

Figures \ref{feh_afe}-\ref{feh_ncfe} show the [X/Fe] 
ratios as a function of [Fe/H] for all sample stars 
described in Section \ref{sec:sample}. As introduced
 in Figure \ref{vphizmax}, the inner and the outer 
halo samples are represented by {\it open} and {\it filled} 
symbols while the colors of the symbols display their 
rotational motions ({\it gray} for $-50<$$V_{\phi}\leqq$150 km s$^{-1}$, 
{\it blue} for $V_{\phi}>$150 km s$^{-1}$ and {\it red} for 
$V_{\phi}\leqq-50$ km s$^{-1}$).
Shaded regions in Figure \ref{feh_afe}-\ref{feh_ncfe} 
show an average and 1$\sigma$ scatter of [X/Fe] for 
the sample stars having typical inner halo kinematics, 
estimated within each metallicity interval with widths of 
$\Delta$[Fe/H]$=0.5$ dex. Here, the 
typical inner halo kinematics is assumed to be
$Z_{\rm max}\leqq5$ kpc and $-50<$$V_{\phi}\leqq$150 km s$^{-1}$
as described in Section \ref{sec:kin}, for which the sample stars are  
represented by the {\it gray open} circles in the figures.

Average [X/Fe] values and their errors ($\sigma /\sqrt{\mathstrut N}$) for 
the inner and the outer halo samples within 
each of four metallicity intervals are summarized in 
Table \ref{abaverage}. In this table, the values are 
shown only for domains whose number of sample stars 
exceeds 1. 

\subsubsection{$\alpha$-elements; Mg, Si, Ca, Ti}

Figure \ref{feh_afe} shows four $\alpha$-elements 
(from {\it top-left} to {\it bottom-right}, Mg, Si, Ca, Ti) 
to iron ratios as a function of [Fe/H].

The most prominent feature in this figure is that a discrepancy 
in the [Mg/Fe] between the inner and the outer halo samples is 
seen in the metallicity interval of $-2<$[Fe/H]$<-1$.
Specifically, the average [Mg/Fe] ratio of the outer halo sample
is lower than the inner one by $0.11$ dex 
as summarized in Table \ref{abaverage}. 
This tendency of the lower [Mg/Fe] ratios is seen among all 
$V_{\phi}$ domains for the outer halo sample.
The lower [Mg/Fe] ratios for the outer halo are not seen among 
more metal-poor stars with [Fe/H]$<-2$. Instead, a large scatter 
in [Mg/Fe] for the outer halo sample, spanning a range in 
[Mg/Fe] of $\sim$ 0.1-0.8 dex, is evident compared to the inner one.
Among stars with [Fe/H]$>-1$, one outer halo star has the 
lowest [Mg/Fe] value.
 Apparent clustering of sample stars at 
$-1<$[Fe/H]$<0$ and [Mg/Fe]$\sim$0.3-0.5
is predominantly occupied by the inner halo sample 
having significant prograde rotation ({\it open blue} circles), which is typical of 
thick-disk stars. 
It is remarkable that these [Mg/Fe] ratios in each [Fe/H] interval for the 
outer halo sample show a clear gradient in [Mg/Fe] with [Fe/H] as opposed to the 
inner halo sample, which shows constant mean [Mg/Fe].
This result confirms the recent $\alpha$ abundance estimate by Z09 
using a subset of the sample in the present work.
 
The [Si/Fe] and [Ca/Fe] ratios tend to be lower for the 
outer halo sample in the 
metallicity range of $-2<$[Fe/H]$<-1$ as seen in the [Mg/Fe]
ratios but by smaller degree.  
In the [Si/Fe]-[Fe/H] plot, the large scatters can be attributed in part to 
the Si lines used in our sample, which are typically weak ($<$30 m{\AA}).
Since no Si lines were detected in more 
metal-poor stars, we could not derive any conclusion on the 
[Si/Fe] behaviors in the metallicity [Fe/H]$<-2$. 
The [Ca/Fe] ratios show a relatively tight correlation with [Fe/H]
compared to the other $\alpha$-elements among both the inner and 
the outer halo samples. Specifically, both samples show decreasing 
trend in [Ca/Fe] with increasing [Fe/H] in all metallicity ranges. 

Difference in the [Ti/Fe]-[Fe/H] relation between the inner 
and the outer halo sample is not clear, except that 
a significant scatter in the [Ti/Fe] ratios for [Fe/H]$<-2$ is evident 
among the outer halo sample. It is also noticeable that 
the decreasing trends of [X/Fe] with increasing [Fe/H] seen in 
Mg, Ca and Si are not evident for the [Ti/Fe] ratios 
for the outer halo sample.

We discuss in 
Section \ref{ab_kin} in detail for this dependence of abundance 
ratios on the orbital parameters.

\subsubsection{Na}

Figure \ref{feh_nafe} shows the [Na/Fe] as a function of
[Fe/H] for the sample stars. 

Na abundances are estimated in this work from Na I doublet lines at 
5682/5688 {\AA} and 6154/6160 {\AA}. 
These lines are typically weak in metal-poor stars ($\lesssim$20 m{\AA}) 
and are less affected by NLTE effects, which are prominent in 
stronger resonance lines \citep{key-39,key-38}. 
NLTE calculations by \citet{key-76} reported that a LTE analysis using
these lines can overestimate Na abundances by 0.05-0.14 dex. 
This amount of correction is comparable to typical abundance errors in this work ($\sim 0.1$ dex).
Thus, we do not apply the NLTE correction to the resulting 
Na abundances. 
 
Derived [Na/Fe] ratios show a large scatter ($\sim$0.15-0.20 dex)
for both of the inner and the outer halo 
samples. The modest difference in the [Na/Fe] ratios 
between the two samples is seen in the metallicity 
interval of $-2<$[Fe/H]$<-1$, such that an average of the [Na/Fe] ratios 
for the outer halo sample ($\mu=-0.31$ dex) is 
lower than the inner one by more than $0.2$ dex (Table \ref{abaverage}). 
This lower average [Na/Fe] ratio for the outer halo sample 
is similar to the value obtained among stars with high rest frame velocity, 
$V_{\rm RF}>300$ km s$^{-1}$, in \citet{key-8}, where the mean [Na/Fe] 
is $-0.34$ for the same metallicity interval. The lower [Na/Fe] was 
also reported for three $\alpha$-poor halo 
stars in \citet{key-16}.

\subsubsection{Fe-peak element; Cr, Mn, Ni, Zn}

Figure \ref{feh_fepfe} shows the abundance ratios for 
Cr, Mn, Ni and Zn as a function of [Fe/H]. For these 
elements, the difference in the abundance ratios 
between the inner and the outer halo samples are not 
clearly seen, except for a modest depression in [Zn/Fe]
for the outer halo sample. 

Specifically, trends in [Cr/Fe] with [Fe/H] 
for the inner and the outer halo samples are 
almost indistinguishable for the metallicity [Fe/H]$>-2$. 
In this metallicity range, 
both of the inner and the outer halo samples 
approximately show the solar value, 0.0 dex, with a small
scatter of 0.1 dex. In the lower metallicity range, 
the outer halo sample shows a slightly 
decreasing trend with decreasing [Fe/H] 
with a relatively large scatter ($0.14$ dex).

The average [Ni/Fe] ratios for the inner and the outer halo 
are indistinguishable.
For both samples, the [Ni/Fe] ratios 
are slightly higher for $-3<$[Fe/H]$<-2$ on average. 

For the [Mn/Fe] ratios, the outer halo 
sample shows an increasing trend with [Fe/H] 
similar to the inner halo sample.
Mn is thought to be a product of explosive 
nucleosynthesis during SNe explosion. 
\citet{key-17} suggest that the trend with [Fe/H] 
can be explained by a higher relative
 contributions from a layer containing Mn 
in metal-rich stars rather than in metal-poor stars.

Many of the outer halo sample show 
[Zn/Fe]$<0.0$ while majority of the inner halo sample
 shows [Zn/Fe]$>0$ in all metallicity range.
Interestingly, lower [Zn/Fe] ratios have been reported for the MW 
satellites. For example, \citet{key-2} reported that 
the [Zn/Fe] ratios determined for Draco, Ursa Minor and Sextans dSphs 
are lower than those of the Galactic halo or disk stars with [Fe/H]$>-2$.   

\subsubsection{Neutron capture element; Ba, Y}

Figure \ref{feh_ncfe} shows [Y/Fe] and [Ba/Fe] ratios as a function 
of [Fe/H] for our inner and the outer halo samples. 
Because most of the inner halo sample with 
[Fe/H]$<-2$ do not have available Y and Ba measurements, the similarity 
and differences between 
the inner and the outer halo samples for [Fe/H]$<-2$ could not be 
investigated in detail in the present work. Nevertheless, 
major features in [Y/Fe]/[Ba/Fe] vs. [Fe/H] seen in the sample of 
 \citet{key-1} without adequate kinematic parameter estimates
are well reproduced with the outer halo sample plotted in 
Figure \ref{feh_ncfe}. Specifically, the [Y/Fe] ratios show 
an increasing trend with [Fe/H] below [Fe/H]$\simeq -1.5$, while it is 
relatively flat for more metal-rich stars. Like Y, Ba for the 
outer halo sample also 
shows an increasing trend with [Fe/H] below [Fe/H]$\simeq -1.5$ which 
reaches below [Ba/Fe]$\simeq -1$ at the lowest metallicity. 
We note that one of two objects having exceptionally high 
[Y/Fe] and [Ba/Fe] ratios is a binary star, BD$+$\timeform{04D} 2466, 
whose anomalous [Ba/Fe] are reported by Z09. 
There is no conclusive evidence of binary nature for another 
Y/Ba-rich star, G 18-24 \citep{key-30}.
An order of $\sim 1000$ scatter observed in \citet{key-1} 
sample in the metallicity interval $-3.2\leq$[Fe/H]$\leq-2.8$ 
was not evident in the present outer halo sample. 
 
\subsubsection{Abundance pattern of the inner and the outer halo stars}

\begin{figure}
  \begin{center}
    \FigureFile(80mm,80mm){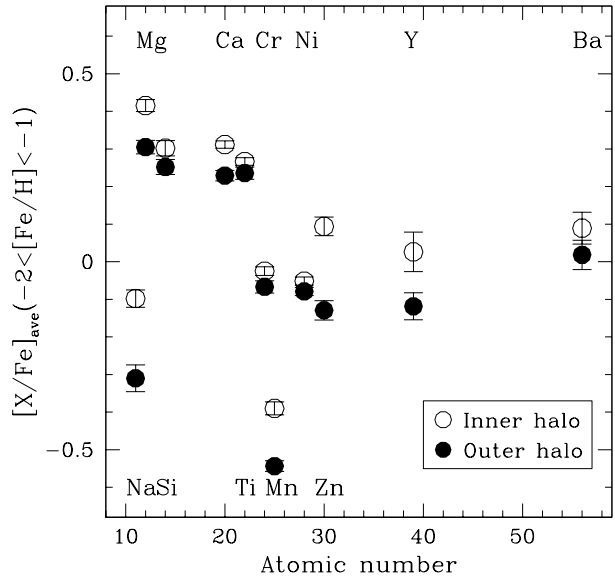}
  \end{center}
  \caption{Abundance pattern for the inner ({\it open} symbols) and the outer 
({\it filled} symbols) halo stars. In 
this plot, average [X/Fe] values ([X/Fe]$_{\rm ave}$) in the metallicity range 
of $-2<$[Fe/H]$<-1$ are plotted. The error bar represents a standard deviation 
divided by a square-root of the number of objects ($\sigma /\sqrt{\mathstrut N}$) 
in each of the inner and the outer halo sample.}
\label{abpt}
\end{figure}

As a summary of our results, Figure \ref{abpt} shows abundance 
patterns for the inner ({\it open} symbols) and the  outer ({\it filled} 
symbols) halo stars in the metallicity range of $-2<$[Fe/H]$<-1$. 
The error bar represents a standard deviation divided by a 
square-root of the number of objects ($\sigma /\sqrt{\mathstrut N}$) in each of the inner 
and the outer halo sample. Since these error bars do not include systematic errors, 
systematic differences between the different analysis methods as mentioned in 
Figure \ref{crosscal} should 
be taken into account when the inner and the outer halo comparison 
is made.

As mentioned in the previous subsections, 
a prominent difference between the inner and the outer halo
stars with $-2<$[Fe/H]$<-1$ 
can be seen in the [Mg/Fe] ratio; an average [Mg/Fe] ratio is 
lower for the outer halo sample by $0.12$ dex. Similarly,  
the [Na/Fe], [Ca/Fe], [Mn/Fe], [Zn/Fe] and [Y/Fe] ratios are 
apparently lower for the outer halo sample.
We note that the inner halo sample is mostly from G03, and the 
consistency of the abundance analysis between G03 and the 
present work is not well confirmed (Section \ref{sec:stellaratms} and 
\ref{sec:crosscal}) The difference between inner and the 
outer halo is, however, supported by the gradient of 
[Mg/Fe], as well as [Ca/Fe], only found for the outer halo sample.
We also note that the [Na/Fe] ratios for our outer halo sample could actually be 
even lower since our LTE abundance analysis could overestimate the [Na/Fe] 
ratios up to $\lesssim 0.15$ dex compared to the NLTE analysis 
performed by G03. For other elements, the inner and the outer halo samples 
show indistinguishable abundance ratios in this metallicity range. 

Our results highlight an necessity of increasing a statistical accuracy 
to conclude on any systematic differences in 
detailed abundance patterns for the 
inner and the outer halo stars. In this analysis, 
systematic errors that possibly arise from various sources 
such as model atmospheres, methods in stellar parameter estimates, atomic line data, 
or a treatment of NLTE effects, may mimic the intrinsic 
scatters on the abundance ratios. 
Thus, it is important to remark that a homogeneous analysis 
technique is essential for studying the current subject, 
as is attempted in the current work.

\subsection{Abundance as a function of kinematic parameters}\label{ab_kin}

\subsubsection{$R_{\rm apo}$-[X/Fe] relations}

\begin{figure}
  \begin{center}
    \FigureFile(160mm,160mm){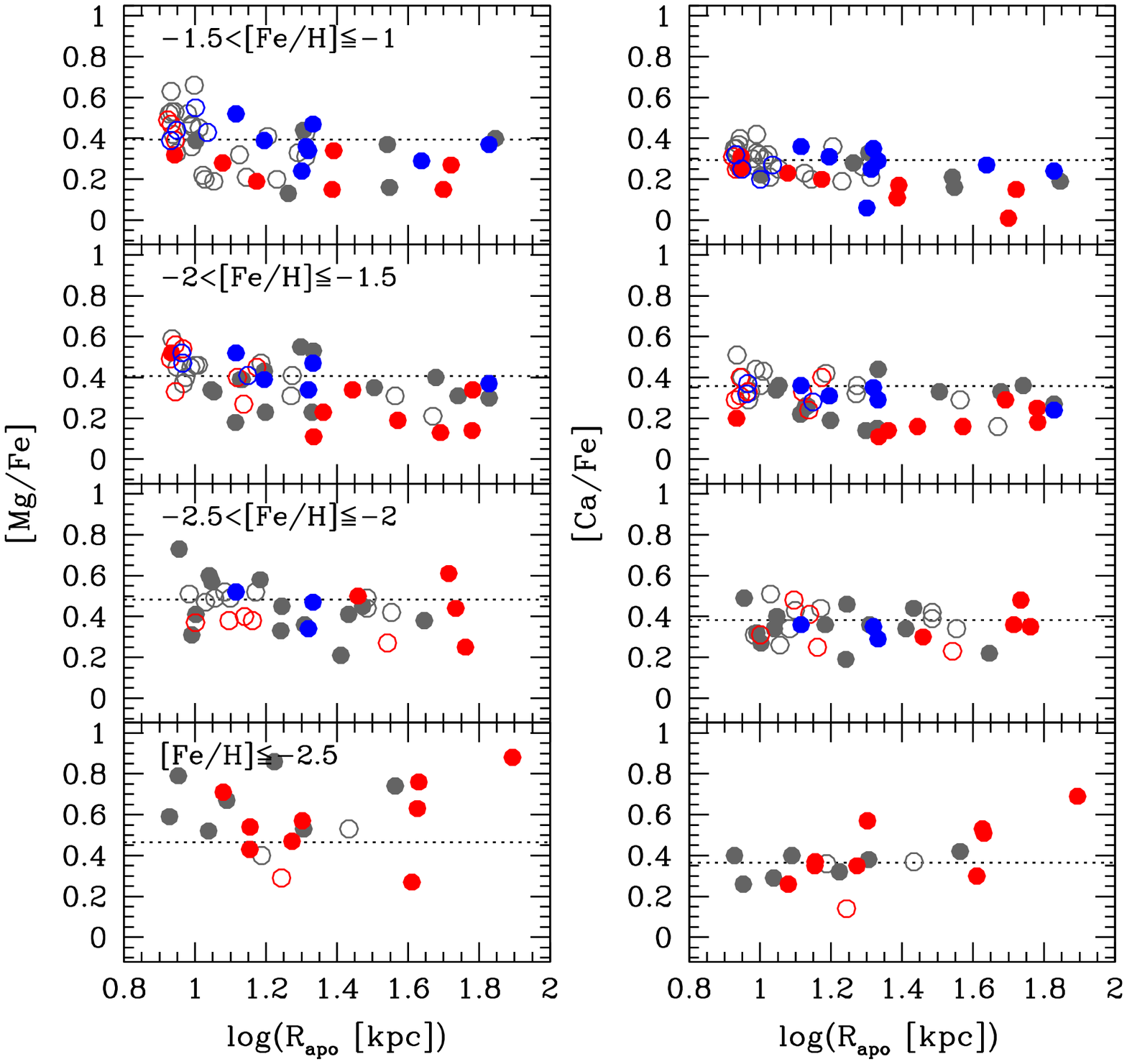}
    \end{center}
  \caption{[Mg/Fe]({\it left} panel) and [Ca/Fe]({\it right} panel) as a function of $R_{\rm apo}$ [kpc]) for different metallicity ranges. The symbols are the same as in previous figures. Doted lines in each panel show the average of the sample of the assumed typical inner halo ($-50<$$V_{\phi}\leqq150$ km s$^{-1}$ and $Z_{\rm max}\leqq5$ kpc) among each metallicity range.}\label{rapo_mg_ca}
\end{figure}

Figure \ref{rapo_mg_ca} shows [Mg/Fe] and [Ca/Fe] 
({\it left} and {\it right} panel, respectively) as a 
function of log($R_{\rm apo}$) for the inner ($Z_{\rm max}\leqq 5$ kpc, 
{\it open} symbols) and the outer ($Z_{\rm max}>5$ kpc,
 {\it filled} symbols) halo samples divided into 
four metallicity intervals; from {\it top} to {\it bottom}, 
$-1.5<$[Fe/H]$\leqq -1.0$, $-2.0<$[Fe/H]$\leqq -1.5$, 
$-2.5<$[Fe/H]$\leqq -2.0$, [Fe/H]$\leqq -2.5$. 
A dotted line in each panel shows an average value of [Mg/Fe] or [Ca/Fe]
for the objects having typical inner halo kinematics 
($-50<$$V_{\phi} \leqq 150$ and $Z_{\rm max}\leqq 5$ kpc). 

The most prominent feature in the [Mg/Fe] vs. $R_{\rm apo}$ plot 
is a difference in the typical [Mg/Fe] values above and below 
$R_{\rm apo}\sim 25$ kpc ($\log R_{\rm apo}\sim 1.4$) for the highest two metallicity intervals. 
Below $R_{\rm apo}\sim 25$ kpc, 
in which the inner halo sample dominates, the sample stars 
show high values of [Mg/Fe] up to $\sim 0.6$
in the both two highest metallicity intervals ([Fe/H]$> -2$).
On the contrary, the sample having larger $R_{\rm apo}$ is mostly 
confined to [Mg/Fe]$\lesssim 0.4$. Such a difference 
above and below $R_{\rm apo}\sim 25$ kpc 
is not evident in the metallicity interval of 
$-2.5<$[Fe/H]$\leqq -2$. Since the sample size of both the inner 
and the outer halo is small, a determination for 
the [Mg/Fe]-$R_{\rm apo}$ trend 
below [Fe/H]$\leqq -2.5$ is unavailable 
for the present sample.    
 
The lower [Mg/Fe] for the larger $R_{\rm apo}$ is reported 
in SB02, in which a weak decline in [$\alpha$/Fe] 
with $R_{\rm apo}$ is observed for the halo stars with 
$-4<$[Fe/H]$<-0.5$. A comparison of their result with
the left panel of Figure \ref{rapo_mg_ca} suggests that 
the decline in the [Mg/Fe] ratio observed for large $R_{\rm apo}$ is
primarily due to the stars with $-2<$[Fe/H]$<-1$. 
Within this metallicity range, the present analysis 
additionally suggests that [Mg/Fe] is rather 
constantly low at $R_{\rm apo}>25$ kpc with 
a negligible gradient. 

A main feature of the [Mg/Fe] ratio described above is also found
 for [Ca/Fe] but by smaller degree; 
consistently lower [Ca/Fe] below $R_{\rm apo}\sim 25$ kpc 
is seen for the sample over the metallicity intervals $-2<$[Fe/H]$<-1$.  
In all metallicity intervals, the scatter in [Ca/Fe] 
is smaller than that of [Mg/Fe].

\subsubsection{$Z_{\rm max}$-[X/Fe] relations}

\begin{figure}
  \begin{center}
    \FigureFile(160mm,160mm){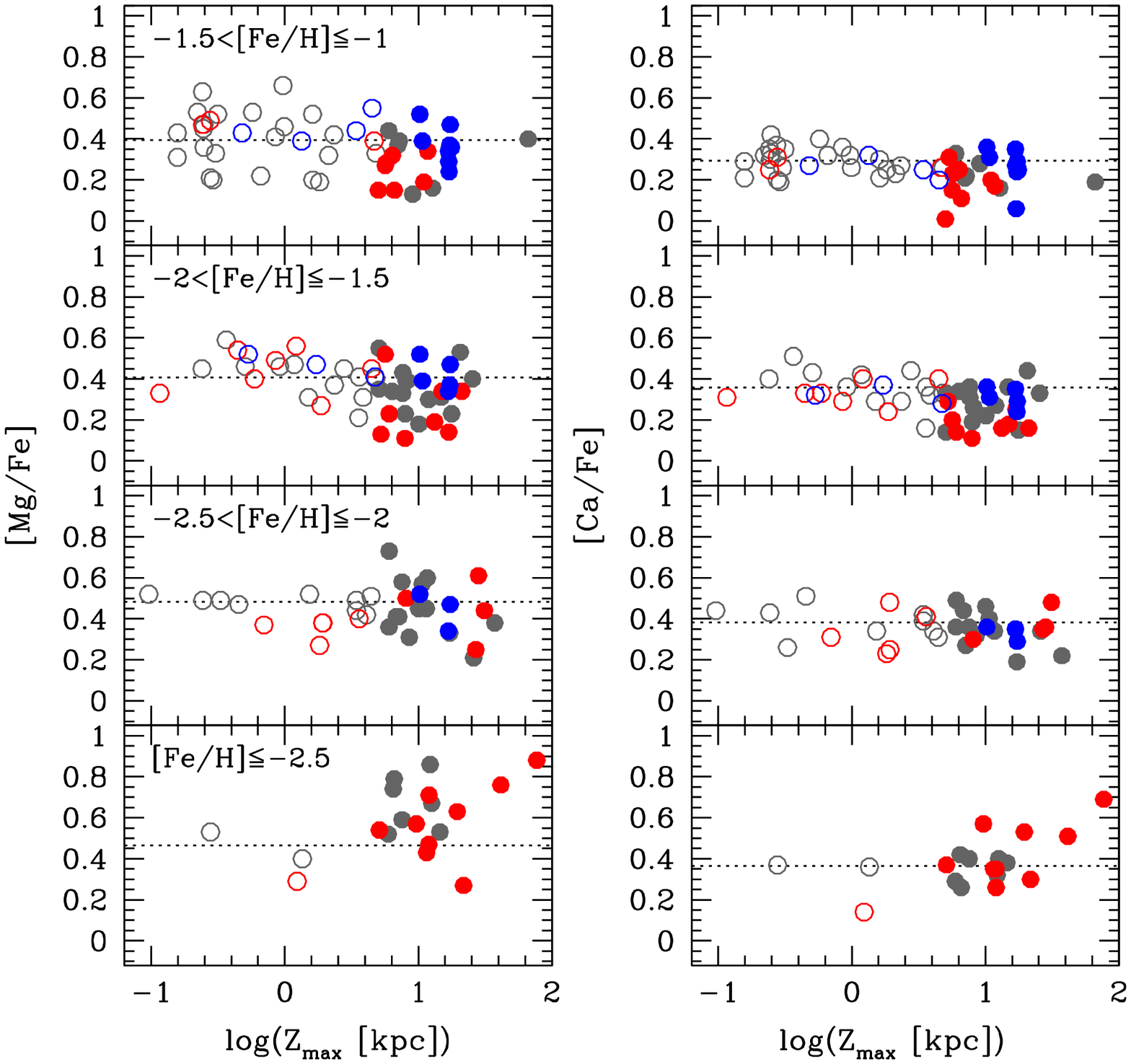}
    \end{center}
  \caption{[Mg/Fe]({\it left} panel) and [Ca/Fe]({\it right} panel) as a function of log($Z_{\rm max}$ [kpc]) for different metallicity ranges. The symbols are the same as in previous figures. Doted lines in each panel show the average of the sample of the assumed typical inner halo ($-50<$$V_{\phi}\leqq150$ km s$^{-1}$ and $Z_{\rm max}\leqq5$ kpc) among each metallicity range.}\label{zmax_mg_ca}
\end{figure}

Figure \ref{zmax_mg_ca} shows [Mg/Fe] and [Ca/Fe] 
({\it left} and {\it right} panel, respectively) as a 
function of log($Z_{\rm max}$) for the inner and the 
outer halo samples with each metallicity interval the same  
as in Figure \ref{rapo_mg_ca}. 
The symbols are the same as in previous 
figures. 

For [Mg/Fe], trends with log($Z_{\rm max}$) are 
significantly different depending on metallicity intervals. 
For stars with $-1.5<$[Fe/H]$\leqq -1$, the inner 
halo stars show a larger scatter in [Mg/Fe] ranging from 
$\sim$0.2 to $\sim$0.7 dex than the outer halo sample for 
which [Mg/Fe] values are mostly below 0.5 dex. 
In the metallicity interval of $-2<$[Fe/H]$\leqq -1.5$, 
an apparent gradient above $Z_{\rm max}\sim 1$ kpc is seen 
among the inner and outer halo samples.
Below $Z_{\rm max}\sim 1$ kpc, the stars with [Mg/Fe]$\gtrsim 0.4$ 
dominate while above $Z_{\rm max}>10$ kpc, the 
average [Mg/Fe] reaches down to $\sim 0.2$ dex.  
In the metallicity interval $-2.5<$[Fe/H]$\leqq-2$,
a large scatter in the [Mg/Fe] ranging from [Mg/Fe]$\sim 0.1$ to
$\sim 0.7$ dex is seen for the outer 
halo sample but not for the inner halo sample. 
The large scatter for the outer halo sample is even more 
prominent in the lowest metallicity [Fe/H]$\leqq-2.5$.

For [Ca/Fe], a modest gradient with log($Z_{\rm max}$) is 
seen in the metallicity interval $-2<$[Fe/H]$\leqq-1.5$ like that 
seen in [Mg/Fe].

\subsection{Dependence on $V_{\phi}$ for [Mg/Fe] and [Ca/Fe]}

\citet{key-35} suggest that the MW outer halo is 
characterized by a net retrograde rotation 
($\langle {\rm V}_{\phi}\rangle<0$ km s$^{-1}$) relative to the 
Galactic rest frame. This result implies that the sample 
stars with a retrograde rotation are more adequate 
representatives of the actual outer halo population rather than 
those with prograde rotation. 

\begin{figure}
  \begin{center}
    \FigureFile(80mm,80mm){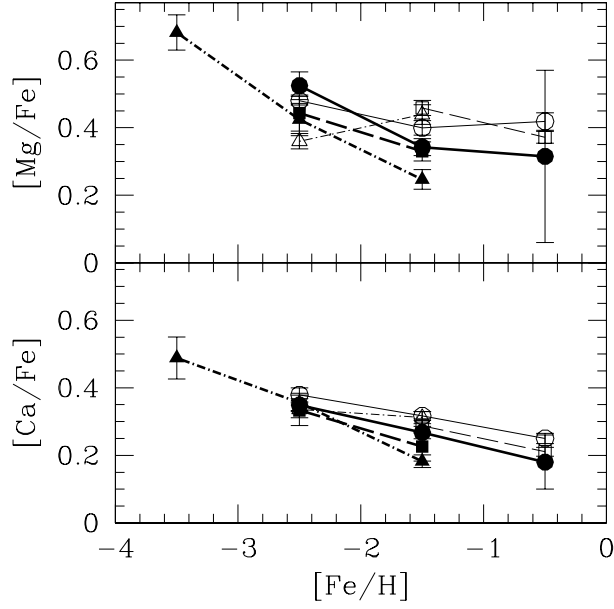}
    \end{center}
  \caption{Average [Mg/Fe] ({\it top}) and [Ca/Fe] ({\it bottom}) at each [Fe/H] bin for different kinematic subgroups. {\it Open} and {\it filled} symbols correspond to the inner and the outer halo sample, respectively. {\it Square}, {\it circle} and {\it triangle} correspond to $V_{\phi}>150$, $-50<$$V_{\phi}\leqq150$ and $V_{\phi}\leqq-50$ km s$^{-1}$, respectively. The error bars represent standard deviations divided by a square root of the number of objects in each [Fe/H] and kinematic domain.}\label{fe_mgca_rot}
\end{figure}

In order to examine possible dependence on $V_{\phi}$ for 
abundance ratios, Figure \ref{fe_mgca_rot} plots 
average [Mg/Fe] ({\it top}) and [Ca/Fe] 
({\it bottom}) for the kinematic subgroups defined by 
$V_{\phi}$ and $Z_{\rm max}$ for each metallicity 
interval. {\it Square, circle} and {\it triangle}  
correspond to $V_{\phi}>150$ km s$^{-1}$, 
$-50<$$V_{\phi}\leqq150$ km s$^{-1}$ and $V_{\phi}\leqq -50$ km s$^{-1}$, 
respectively. 
The {\it open} and the {\it filled} symbols represent 
the sample with $Z_{\rm max}\leqq 5$ kpc and $Z_{\rm max}>5$ kpc 
, corresponding to the 
inner and the outer halo sample, respectively. 
The error bars show standard deviations of the [Mg/Fe] ratios in 
each metallicity interval and the kinematic subgroup
divided by a square-root of a number of the sample stars 
averaged over. 
 
In the metallicity interval of $-2<$[Fe/H]$<-1$, 
it is noticeable that the average 
[Mg/Fe] ratio for the outer halo sample with significant retrograde 
rotation, $V_{\phi}\leqq -50$ km s$^{-1}$, is slightly lower than the 
other two $V_{\phi}$ subgroups. This can be interpreted as 
the objects with significant retrograde rotation are more likely 
belonging to the outer halo, which is thought to 
be formed with a different mechanism from the inner halo \citep{key-35}.
The [Ca/Fe] ratios are almost indistinguishable between the three $V_{\phi}$ domains. 
In generally, the present sample with $-2<$[Fe/H]$<-1$ does not show 
significant difference in the [Mg/Fe] and [Ca/Fe] ratios 
between the different $V_{\phi}$ ranges, 
while the abundance difference is more evident between 
the inner and the outer halos as defined by 
below and above $Z_{\rm max}=5$ kpc in this study.

\subsection{Kinematics of the low-[Mg/Fe] stars }

\begin{figure}
      \begin{center}
           \FigureFile(155mm,155mm){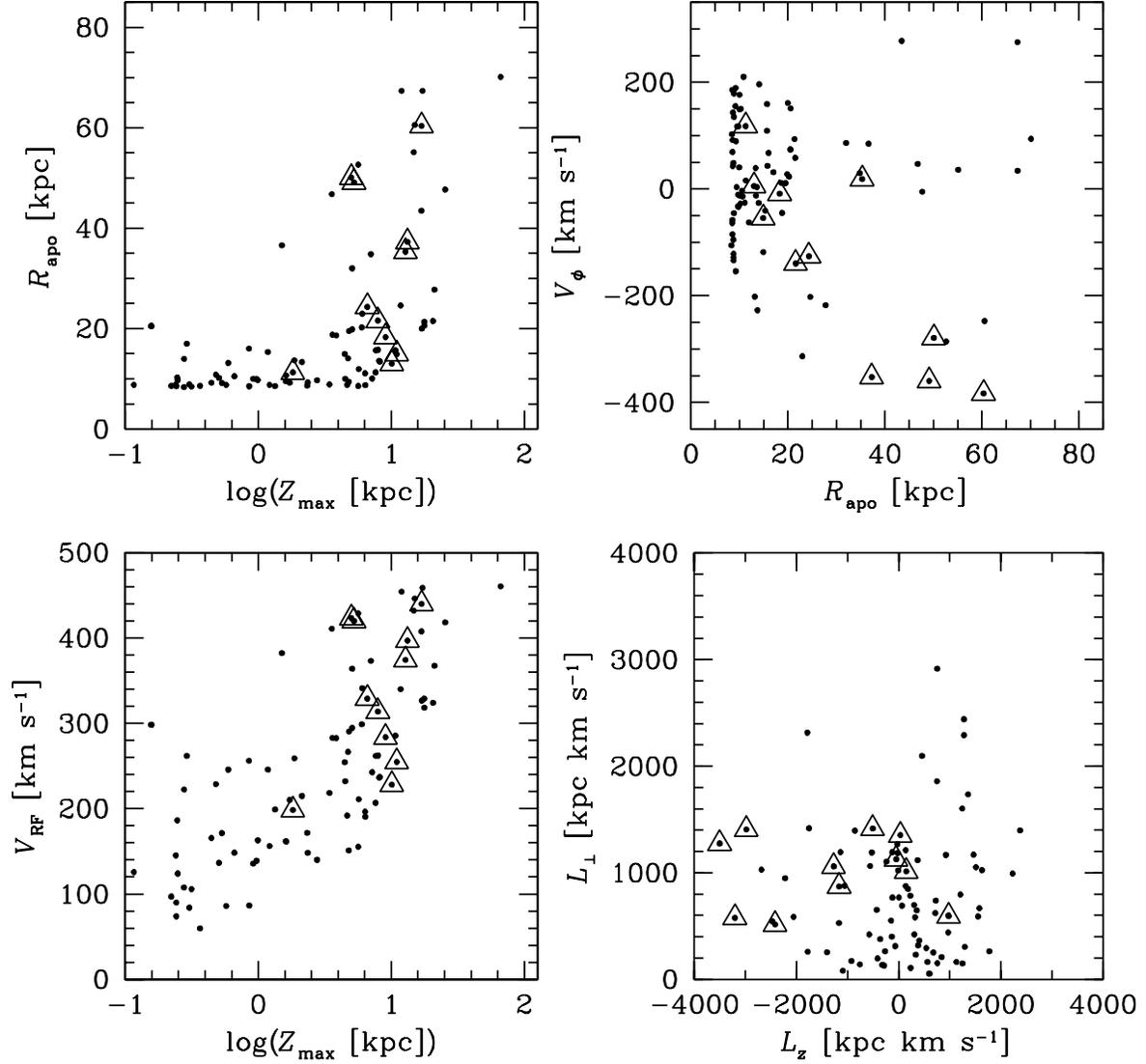}
	   \end{center}
  \caption{Distribution of the sample stars in the planes defined in the orbital
parameters. The sample stars with $-2<$[Fe/H]$<-1$ are plotted. 
{\it top-left}: $Z_{\rm max}$-$R_{\rm apo}$, {\it top-right}: $R_{\rm apo}$-$V_{\phi}$, 
{\it bottom-left}: $Z_{\rm max}$-$V_{\rm RF}$, {\it bottom-right}: 
$L_{z}$-$L_{\perp}$ . The stars with [Mg/Fe]$<0.2$ are highlighted 
with {\it open triangles}. }
  \label{kin_cor_ab}
\end{figure}

Four panels of Figure \ref{kin_cor_ab} show correlations
 of the orbital parameters
similar to Figure \ref{kin_cor}  but now stars with [Mg/Fe]$<0.2$ 
(hereafter ``low-[Mg/Fe] stars'') 
are highlighted with open triangles. Only sample stars 
with $-2\leqq$[Fe/H]$\leqq -1$
are plotted in the figure. 

In the $Z_{\rm max}$-$R_{\rm apo}$ plane ({\it top-left} panel), 
almost all low-[Mg/Fe] stars
reside in the region, $5<$$Z_{\rm max}<25$ kpc ($0.7<\log Z_{\rm max}<1.4$),
except one star at $Z_{\rm max}<5$ kpc. The low-[Mg/Fe] stars
distribute over a wide range in $R_{\rm apo}$ 
such that $R_{\rm apo}\sim 10-65$ kpc. 
The {\it top-right} panel shows a 
distribution of the sample stars in a $R_{\rm apo}$-
$V_{\phi}$ plane. In the range $R_{\rm apo}<20$ kpc, 
most low-[Mg/Fe] stars reside around $V_{\phi}\sim 0 \pm 100$ km s$^{-1}$. 
At higher $R_{\rm apo}$, the low-[Mg/Fe] stars tend to have a retrograde 
orbit. The {\it bottom-left} panel 
shows the distribution of the sample stars in a 
$Z_{\rm max}$-$V_{\rm RF}$ plane. \citet{key-8} reported that, 
based on chemical abundance analysis for a 
large sample of $>70$ disk and halo stars, [Mg/Fe] ratios are 
lower by $\sim 0.2$ dex for the objects 
with $V_{\rm RF}>300$ km s$^{-1}$. 
The lower [Mg/Fe] ratios for $V_{\rm RF}>300$ km s$^{-1}$ stars 
are also seen in the $Z_{\rm max}$-$V_{\rm RF}$ plot in Figure \ref{kin_cor_ab}. 
Specifically, 27\% of sample stars 
with $V_{\rm RF}>300$ km s$^{-1}$ are low-[Mg/Fe] stars 
while this fraction drops for lower $V_{\rm RF}$ stars: 
13\% and 0 \% for $150<$$V_{\rm RF}\leqq300$ km s$^{-1}$
and $V_{\rm RF}\leqq150$ km s$^{-1}$, respectively.

\begin{figure}
      \begin{center}
           \FigureFile(80mm,80mm){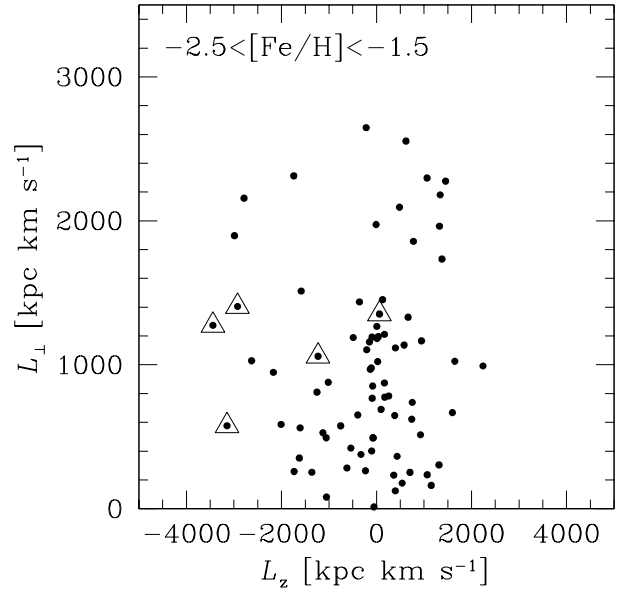}
	   \end{center}
  \caption{Same as {\it bottom-right} panel of Figure \ref{kin_cor_ab} 
but now for metallicity interval of $-2.5<$[Fe/H]$<-1.5$}
  \label{lz_ltan}
\end{figure}

In the angular momentum $L_{z}$-$L_{\perp}$ space ({\it bottom-
right} panel), low-[Mg/Fe] stars predominantly 
reside at $L_{\perp}\gtrsim 600$ kpc km s$^{-1}$. In the
$L_{z}$-$L_{\perp}$ plane, \citet{key-26} identified a kinematic 
substructure (H99 stream), for which several other studies confirmed 
the presence of a similar structure \citep{key-4,key-27}.
Recently, \citet{key-53} suggest that, for 21 candidate members 
of the H99 stream, 
the [Fe/H] distribution peaks at [Fe/H]$\approx-2.0$. 
Figure \ref{lz_ltan} is the same as the {\it bottom-right} 
panel of Figure \ref{kin_cor_ab}, except that the metallicity range 
is now set to $-2.5<$[Fe/H]$<-1.5$ which is comparable to 
the suggested metallicity distribution of the H99 stream.
In a region containing the reported location of the H99 stream in 
the $L_{z}$-$L_{\perp}$ plane ($L_{\perp}\approx 2000$ 
and $L_{z}\approx500$-$1500$ kpc km s$^{-1}$), no low-[Mg/Fe] stars 
are included and span [Mg/Fe] ratios of $\sim0.23-0.52$.
Whether these stars represent abundances of this kinematic 
substructure is quite uncertain in the present sample 
because of the small sample size and the incomplete sample selection. 
Future systematic observations for robust kinematics and chemical 
abundances of halo stars including fainter objects 
may be required to address whether such kinematic 
substructures have distinct chemical abundances 
compared to stars in the smooth halo component.

\section{Discussion}
\label{sec:disc}

\subsection{Comparison with recent studies of abundances and kinematics for the halo stars}

Previous studies for the possible dependence of detailed chemical 
abundances of solar-neighborhood stars on their kinematics have provided
important implications for early formation of the MW 
(e.g., \cite{key-68, key-69, key-8}; SB02; \cite{key-9}; \cite{key-36}). 
In our work, 
a primary improvement compared with previous studies is the inclusion of 
a large number of stars with $Z_{\rm max}>5$ kpc. This improvement has 
the following advantages that 1) abundance-kinematics correlation 
can be examined by finely
dividing the sample into small metallicity intervals and 2) systematic 
abundance difference between the assumed inner/outer 
components can be investigated covering a larger number of 
the outer halo sample, not restricted to a few stars with extreme 
orbital properties. 

SB02 conducted an abundance analysis of 
high-resolution and high signal-to-noise spectra
for halo stars, most of which have extreme kinematic
properties such as a large $R_{\rm apo}$, a 
large $Z_{\rm max}$ or an extreme 
retrograde orbit ($V_{\phi}<0$ km s$^{-1}$). 
They showed that the [$\alpha$/Fe] ratios
are lower for large $R_{\rm apo}$
stars at a given metallicity without clear connection 
with both of $Z_{\rm max}$ and $V_{\phi}$. Their result for 
the dependence of the [$\alpha$/Fe] ratios on $R_{\rm apo}$ 
and $V_{\phi}$ is indeed reproduced in this work with a  
much larger number of sample in each metallicity interval. 
Specifically, our results suggest that 
stars with $R_{\rm apo}>25$ kpc have lower [Mg/Fe] ratios, 
which are a main contributor to [$\alpha$/Fe]-
kinematics correlation, 
primarily in the metallicity [Fe/H]$>-2$ (Figure \ref{rapo_mg_ca}) without 
showing a clear dependence on $V_{\phi}$ (Figure \ref{fe_mgca_rot}). 
Our result also suggests that 
[Mg/Fe] ratios are lower for large $Z_{\rm max}$ stars
as well, as found by increasing the number of the sample stars 
with $Z_{\rm max}>5$ 
kpc (Figure \ref{zmax_mg_ca}). This feature additionally highlights
the importance of examining abundance-kinematics correlations within 
each metallicity interval rather than including all stars with [Fe/H]$<-1$, 
since the correlation could be significantly different 
depending on a given metallicity. 

Another important improvement in the present work is that the 
sample of all 57 halo stars, most of which have outer halo kinematics, 
were observed with similar settings and analyzed in a homogeneous manner. 
The homogeneous analysis allows us the examination of intrinsic scatter of 
the outer halo sample without been significantly affected by systematic errors.
Recently, \citet{key-36} used a large sample of abundances and kinematics 
assembled from literature to investigate any 
systematic trends in abundances with their kinematic properties.  
Although more stringent criteria for the inner and the outer halo stars 
were employed, each domain largely overlaps with the simple classification 
employed in the present work. A consistent result is obtained 
in both works that the [Mg/Fe] in the intermediate metallicity range 
is lower for the outer halo stars than for the inner halo stars. 
A larger scatter in the [Ni/Fe] for the outer halo
than the inner halo reported in \citet{key-36} is not reproduced 
in the present work. This feature implies that the reported large 
scatter in [Ni/Fe] may be partly attributed to the systematic 
errors among the different abundance analyses. Nevertheless, the present work 
does not properly cover the metallicities
below [Fe/H]$<-3$ yet. More investigation is needed to conclude on the 
inner/outer difference in a scatter of [Ni/Fe] at lowest metallicity.

\subsection{The inner/outer halo boundary in terms of the abundance ratios}
\label{sec:boundary}

\begin{figure}
       \begin{center}
         \FigureFile(155mm,155mm){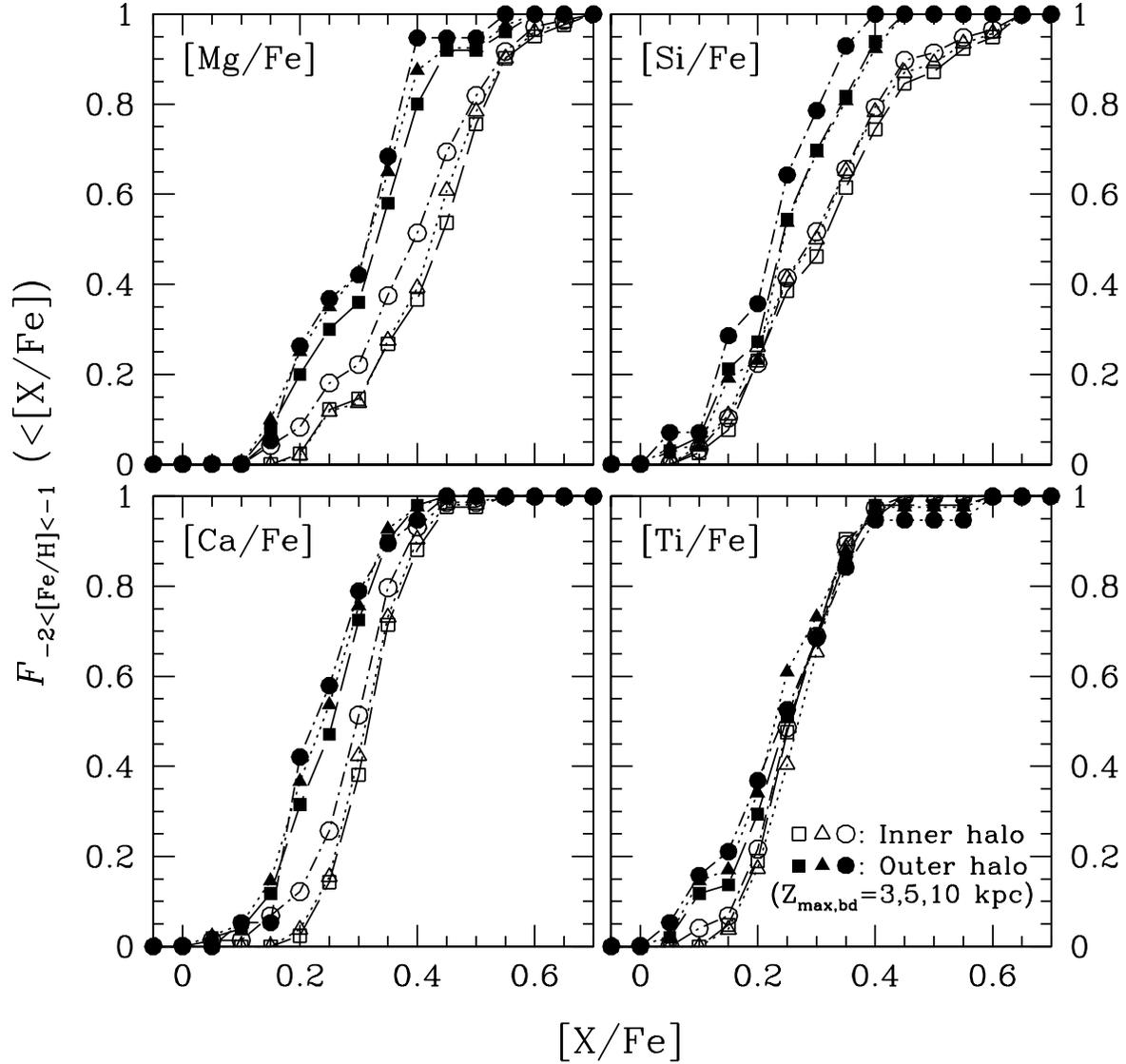}
         \end{center}
   \caption{Cumulative distribution of [X/Fe] for the sample stars with $-2<$[Fe/H]$<-1$ adopting various $Z_{\rm max}$ boundaries ($Z_{\rm max, bd}=$3, 5, 10 kpc) to separate the inner and the outer halos. {\it Squares}, {\it triangles}, {\it circles} show distributions by setting the boundary as $Z_{\rm max, bd}=$3, 5, 10 kpc, respectively. {\it Open} and {\it filled} symbols show the corresponding distributions for the inner and the outer halo, respectively. }
  \label{cum}
\end{figure}

As can be seen in the previous sections, setting the conventional 
inner/outer halo boundary as being $Z_{\rm max}=5$ kpc results in systematic 
difference in [Mg/Fe], and also in [Si/Fe]
and [Ca/Fe] with a smaller degree, between 
these two halo components in the metallicity 
range of $-2<$[Fe/H]$<-1$. 
Validity of setting $Z_{\rm max}=5$ kpc as the likely 
inner/outer halo boundary is 
examined by applying the different boundary values of $Z_{\rm max}=$3 and 10 kpc as well. 
In order to illustrate this examination,  Figure \ref{cum} 
shows differences in 
cumulative [Mg/Fe], [Si/Fe], [Ca/Fe] and [Ti/Fe] distributions for the assumed 
inner/outer halo by setting different 
$Z_{\rm max}$ boundaries (3, 5, 10 kpc). Only stars with the metallicity range of
 $-2<$[Fe/H]$<-1$ are included in this analysis.
Table \ref{ks} summarizes  
Kolmogorov-Smirnoff (K-S) probabilities for the inner/outer 
halo division in terms of the four $\alpha$ elements (column 2-5) 
with the adopted $Z_{\rm max}$ boundaries (column 1). 
For Mg, whose discrepancy between the inner/outer halo is prominent, 
the change in the boundary within $Z_{\rm max}=3$-$10$ kpc does not largely 
alter the conclusion. More specifically, Table \ref{ks} shows that,
regardless of the different choices of 
the boundaries, K-S tests for the assumed inner/outer
 halo [Mg/Fe] distributions disprove the null hypothesis that the two 
distributions are drawn from the same distribution at P$_{\rm KS}<0.01$.
For Ca, discrepancy such that P$_{\rm KS}<0.01$ is observed if 
the boundary of $Z_{\rm max}=$3 or 5 kpc is set. On the other hand, 
for Si and Ti, a K-S test cannot disprove 
the null hypothesis at P$_{\rm KS}<0.01$for 
any choice of the boundaries. Especially, 
the assumed inner/outer halo are indistinguishable in terms of [Ti/Fe]. 

These results suggest that, regarding the present sample, the inner/outer halo 
transitions in a vertical direction are rather continuous in terms of Mg, Si, 
Ca and Ti abundance ratios in the range of $Z_{\rm max}=3$-$10$ kpc. 
A sharp boundary within this vertical range is not prominent in 
the present analysis. 
It can be noticed, however, that for Mg, the choice of the boundary as 
$Z_{\rm max}=10$ kpc results in an inclusion of a large number of low [Mg/Fe] 
stars in the inner domain. This feature suggests that 
the low [Mg/Fe] stars progressively contribute to the stellar halo at 
$Z_{\rm max}>5$ kpc. Therefore, the present analysis implies 
that the properties 
(IMFs, star-formation timescale, etc.) of the progenitor system, either 
{\it in situ} or accreted dwarf galaxies, that contribute to build up 
the stellar halo, start to change at a vertical distance less than 10 kpc. 

Whether there exists a boundary above which a spatial distribution, 
kinematics and chemical abundances are distinct from the inner 
halo is still unclear. 
From a large data set of positions, distances and full 
space motions for the solar-neighborhood halo stars, \citet{key-35} 
report an increasing contribution from a stellar component with 
 lower peak metallicity ([Fe/H]$\sim -$2.2) and a net retrograde 
rotation as increasing $Z_{\rm max}$ above 5 kpc. 
This change in characteristic metallicity and rotational 
motion supports the suggestion obtained in this work 
from the elemental abundances that  
dominant progenitors may be different 
between the inner and the outer halos.
It is, however, uncertain in both works whether a 
transition from the inner to the outer halo is sharp 
or continuous in terms of metallicity, 
rotational motion and the elemental abundances. 
This issue should be addressed with the expanded 
homogeneous data set of full phase-space information as well as 
elemental abundances for fainter objects not 
restricted to the local volume.

It can also be seen from Figure \ref{cum} that a degree 
of the inner/outer halo distinction varies with elements. 
Mg, Si, Ca and Ti, are conventionally referred to as 
$\alpha$-elements because these are thought to be all produced 
through similar mechanisms (captures of $\alpha$ particles). 
However, the present analysis shows that the degree to which 
the assumed inner/outer halo are different is large for Mg 
but negligible for Ti. This discrepancy could partly be due to  
different production mechanisms for Mg and Ti \citep{key-70}, e.g., 
 Mg is mostly produced with 
hydrostatic burning while dominant Ti is produced during supernovae
explosions. 

\subsection{Comparison with the nearby dSphs; possible progenitors of the outer halo stars}

We now consider the properties of the possible progenitor systems
of the outer halo, whose candidate members show lower [Mg/Fe]
than those of the inner halo in this work.
These progenitor systems are thought to contribute
 to a large fraction of the MW halo according to the 
currently standard $\Lambda$CDM model.
 Although it is now broadly believed  
that the accretions of smaller progenitor systems have played an important 
role in making up the MW stellar halo, the properties of the progenitor systems 
such as a typical size, mass, stellar/gas content, star formation histories 
are still controversial.

One probable candidate as a dominant 
progenitor system of the MW halo is a system similar 
to the dSphs currently orbiting the MW halo. In addition to $\sim$10 dSphs 
previously known (sometimes called ``classical" dSph), $\sim$12 dSphs, 
most of which are fainter than $\sim$10$^{5}$L$_{\odot}$, have been newly 
discovered through the SDSS \citep{key-41, key-42}. 
In the following subsections, 
the possibility that systems similar to these dSphs 
could be the dominant progenitor of 
the MW halo by examining similarity in chemical 
abundance patterns of constituent stars in both systems.

\subsubsection{Abundance overlap with ``Classical'' dSphs}

Abundance analyses for red giant stars belonging 
to the relatively bright MW satellites, 
Ursa Minor, Sculptor, Draco, Sextans, Carina, Fornax and  
Leo I, are available from high-resolution spectroscopy 
(e.g., \cite{key-2, key-12, key-74, key-82, key-13, key-45}). 
A comparison of a detailed abundance pattern for our outer halo sample 
with these satellites, so called ``Classical'' dSphs, 
provides implications about whether any of the outer halo 
sample could have once belonged to systems similar to these dSphs. 
It was reported 
that the chemical abundance patterns for these dSphs are distinct 
compared to the bulk of the Galactic stars at a given metallicity range
in that the [$\alpha$/Fe] ratios are lower for the dSphs \citep{key-15}. 
Slightly lower [Zn/Fe] and [Y/Fe] for the 
dSphs than the halo stars is also reported \citep{key-12}.

\begin{figure}
\begin{center}
  \FigureFile(155mm,155mm){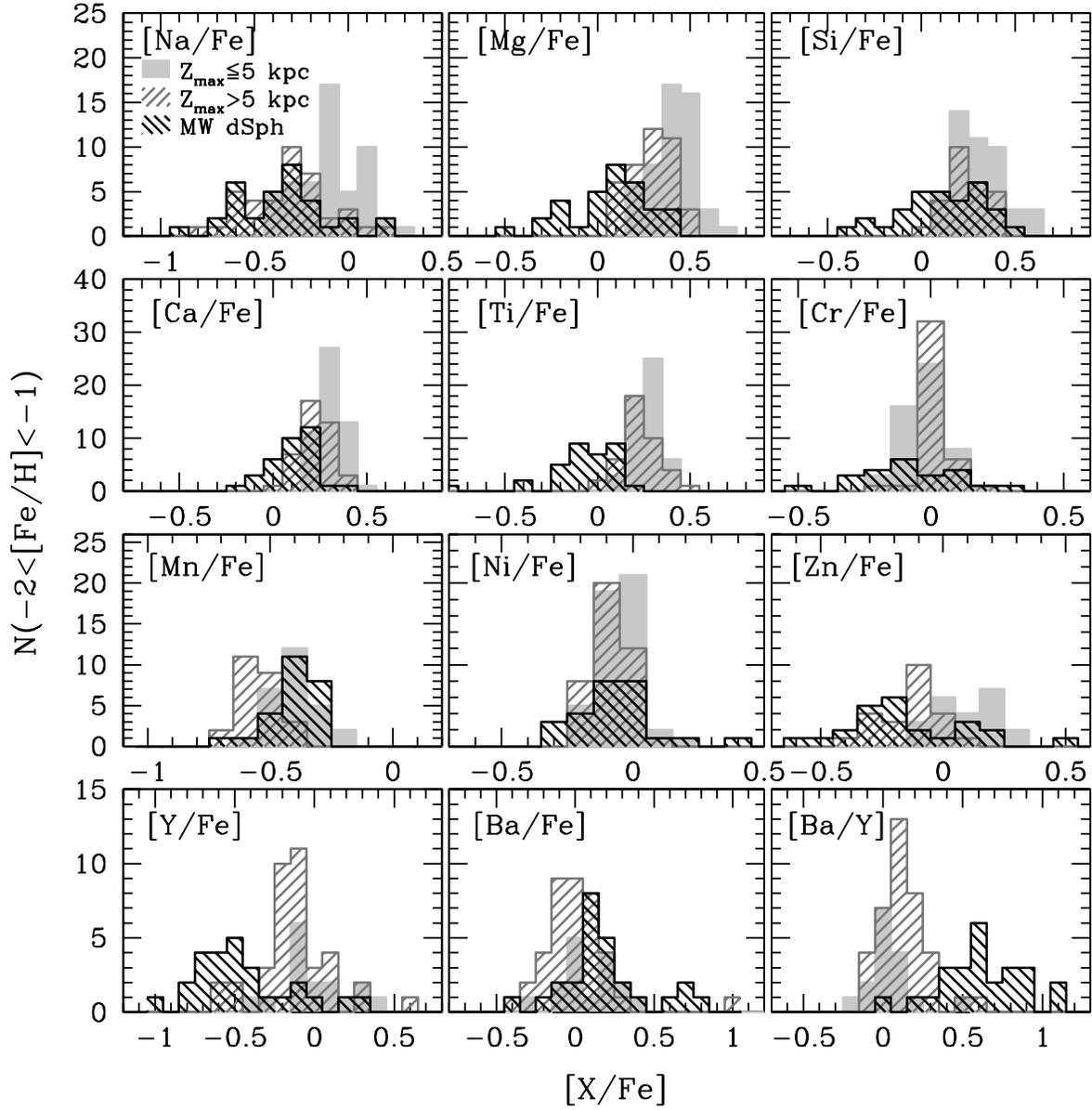}
  \end{center}
  \caption{Distribution of the abundance ratios for the inner and the outer halo samples 
({\it gray-solid} and {\it gray-shaded} histogram, respectively) and the MW dSphs 
({\it black-shaded} histogram) in the metallicity interval $-2<$[Fe/H]$<-1$.  
The abundance data for the dSphs were taken from \citet{key-2}, \citet{key-12}, \citet{key-82} and \citet{key-13}.}
  \label{dsph}
\end{figure}

Figure \ref{dsph} compares the distributions of the [X/Fe] abundance ratios 
determined for the MW dSphs by \citet{key-2, key-12, key-82} and \citet{key-13} 
({\it black shaded} histogram) with those of our inner and outer halo 
samples ({\it gray solid} and {\it gray shaded} histogram, respectively) 
in the MW halo. The metallicity range is restricted to $-2<$[Fe/H]$<-1$, 
where the observed discrepancy in [$\alpha$/Fe] ratios between the dSphs 
and the MW halo stars is largest. For some elements, similar peaks of the 
distributions are evident for the dSphs and the outer halo sample, 
that are distinct from the inner halo sample. 
In particular, the [Mg/Fe] ratio, 
which is shown in the middle panel of the top row, 
is peaked at $\sim$0.3
for the dSphs and the outer halo sample, while that of the 
inner halo sample is peaked at $\sim$0.4-0.5. According to
chemical evolution models for dSphs, the lower [Mg/Fe] ratio is partly
interpreted by a lower star formation rate in dSphs which leads
to a significant contribution of Fe to the system from delayed 
enrichment of Type Ia SNe (e.g., \cite{key-73}). 
Therefore, if the initial mass function (IMF) is the same for 
any systems and time-independent, contributions of Fe from 
SN Ia are needed to reproduce the [Mg/Fe] ratios observed in 
dSphs and the outer halo sample. 
Although the crude assumptions about the 
universality of the IMF should be ultimately examined, it can be 
inferred that possible 
progenitor systems for the outer halo are similar to the dSphs 
in terms of its lower star formation rate.  
Another remarkable feature in a comparison of the [Mg/Fe] distributions 
is that the [Mg/Fe] ratios of $<-0.1$ observed in the dSphs are not observed 
for both of the inner and outer halo samples. The extremely 
low values of the [Mg/Fe] ratios are observed in Draco, Sextans, Carina and 
Leo I \citep{key-2, key-12}. A few Galactic halo stars have 
also been reported to have [Mg/Fe]$<-0.1$ \citep{key-16}, 
similar to the values 
observed in the dSphs. Whether such low [Mg/Fe] ratios are extremely 
rare within the MW halo or not cannot be concluded in the present 
analysis because of the limitation of the number of sample 
($\sim$60 for $Z_{\rm max}>5$ kpc) and the spatial extent 
(within $\sim$2 kpc from the Sun).

The [Na/Fe] and [Zn/Fe] ratios distributions are peaked at slightly lower 
value for the outer halo sample 
than the inner halo sample. 
For the [Na/Fe] ratios, the distribution for the outer halo 
and the dSphs have similar peaks at $\sim-0.3$ dex. 
For the [Zn/Fe] ratios, although the distribution for the outer halo 
partly overlap with that for dSphs, these distribution 
is significantly different in that the distribution for 
dSphs is more broad ranging from $\sim$$-0.6$ to 0.5 
while that for the outer halo spans relatively narrow range. 
Models of nucleosynthesis products in the Type II SNe 
suggest that yields of Zn are determined by various 
factors such as an explosion energy, a progenitor mass, 
a mass cut or a geometry of the explosion \citep{key-17}. 
Which factor is mostly responsible for the [Zn/Fe] of the 
stars with $-2<$[Fe/H]$<-1$, producing global abundance 
difference within the outer halo progenitors, remains unclear.
We note that the measurement quality of dSph stars is not as good as that of 
halo stars, and that might be one of the reasons for the large scatter.
Also, as noted in Section \ref{sec:crosscal}, the [Na/Fe] and the 
[Zn/Fe] ratios are systematically offset by $\sim 0.1-0.2$ dex 
in our outer halo sample, which implies that an actual
difference between the inner and the outer halo is much smaller 
than that can be seen in Figure \ref{dsph} or in Table \ref{abaverage}, if any. 

For the neutron capture elements, Y and Ba, 
the distributions for both of the outer and the inner halo samples 
are significantly different from that for the dSphs. 
Specifically, for the field halo sample, the distributions of [Y/Fe] 
are peaked at higher values while those of [Ba/Fe] are peaked 
at lower values than for the dSphs. As a result, a 
significant discrepancy in distributions of [Ba/Y] ratio between the 
field halo samples and the dSphs is evident as shown in the lower right 
panel of Figure \ref{dsph}.

For lower metallicity ranges, \citet{key-45}
analyzed chemical abundances of extremely metal-poor ([Fe/H]$<-2.5$)
stars in the Sextans dSphs. 
They showed that five in six sample stars show Mg/Fe comparable to 
the solar value, which is significantly lower than the typical Mg/Fe for 
the Galactic halo stars with similar metallicity. 
As shown in Figure \ref{feh_afe}, none of such low Mg/Fe stars are found among
 the outer halo sample with [Fe/H]$<-2.5$, such that [Mg/Fe]$\sim 0.4$-$0.5$ 
on average. This comparison for Sextans dSphs and the MW outer halo 
imply that both systems had experienced distinct chemical 
enrichment history even at earlier epoch when extremely metal-poor 
stars were formed. Further abundance information for 
many stars belonging 
to other dSphs as well as the MW halo stars with various kinematics 
are necessary to conclude on a systematic abundance similarity/difference 
between these systems.

\subsubsection{Comparison with ultra-faint dSphs}

Ultra-faint dSphs recently discovered in SDSS data are a few orders of magnitudes 
fainter than ``Classical'' dSphs (e.g., \cite{key-41}). Metallicities 
of these objects were measured to be [Fe/H]$\sim -3$, which is comparable to  
a low metallicity tail of the MW halo metallicity distribution \citep{key-43}.
Detailed chemical abundances have been measured for several stars belonging 
to these objects \citep{key-14,key-44}. 

\citet{key-14} reported the chemical abundance of two red giants in the 
recently discovered ultra-faint dSph, Hercules, 
which are highly unusual in the following points; 
1) these stars are strongly deficient in Ba ([Ba/Fe]$<-2$) and 
2) Mg is highly enhanced compared to Ca. The derived abundances 
are consistent with the ejecta of massive $\sim 30$ $M_{\odot}$ Type II 
SNe. From these results, it is suggested that these stars were 
formed with ejecta of a few massive stars as a result of 
an incomplete mixing of elements within the star forming clouds. 
Some of the outer halo stars with comparable metallicity 
($-3<$[Fe/H]$<-2$) show similar enhancement in Mg relative to Ca. 
On the other hand, no stars are found to be in this level of depression in Ba.
If a mixing of nucleosynthesis products contained in individual SNe ejecta 
is inefficient in ultra-faint dSphs, a large scatter in abundances of 
heavy elements is expected. It remains unclear whether the 
system like ultra-faint dSphs contributed to the stellar halo until
the expected scatters in the abundances are constrained 
using a larger number of abundance data for stars
 belonging to such faint stellar systems.

\subsection{Implications for the halo formation }

In the previous sections, we have shown our main result 
that the inner and the outer 
halo samples display different trend in the 
[Mg/Fe]-[Fe/H] relation in the metallicity range of 
$-2<$[Fe/H]$<-1$, as found by significantly increasing the 
number of the sample stars
with $Z_{\rm max}>5$ kpc. Additionally, in this 
metallicity range, the outer 
halo sample displays the overlapping [Mg/Fe] distribution
with the MW dwarf satellites. These results suggest that 
the MW halo is not homogeneously enriched with Mg-rich 
ejecta of massive Type II SNe, associated with rapid 
formation of the entire stellar halo which is expected to yield high [Mg/Fe]
ratios (e.g., \cite{key-70}). Instead, the results support the suggestion 
that the inner and the outer halos may have formed 
with different mechanisms as examined by kinematics and 
metallicity of the halo stars (e.g., \cite{key-35}). 
Comparisons with the chemical evolution models specifically 
suggest that dominant progenitors, either {\it in situ} 
or accreted systems, of the inner and the outer halos 
experienced different chemical enrichment history.    
 
Possible progenitors of the low-[Mg/Fe] outer halo stars
are systems in which star formation had proceeded 
at a slow rate and/or accreted relatively recent time. 
The recent accretion allows the progenitor system to dominate 
with iron-rich materials due to 
 delayed enrichment with Type Ia SNe (e.g., \cite{key-71}).  
This scenario requires a longer timescale for the 
outer halo formation which is consistent with the 
possible age spreads for the outer halo globular 
clusters in the MW using metallicities and horizontal-branch 
morphologies as a tracer \citep{key-47, key-46}.
Indeed, likely remnants of the later accretions
were identified as spatially coherent substructures in the MW 
stellar halo (e.g., \cite{key-23}). 
Nevertheless,  Figure \ref{dsph} also shows that the [Mg/Fe] 
distributions for neither of the inner nor the outer halo can  
reproduce the lowest [Mg/Fe] tail ([Mg/Fe]$<0$) of the satellites. 
This result suggests that the progenitor system 
would not completely resemble with the presently surviving 
MW satellites in terms of chemical abundances, 
thereby star-formation history. As suggested in the 
cosmological simulation of \citet{key-49}, the outer halo 
progenitors could have been disrupted through tidal 
interactions with the MW halo, even though they may have 
accreted relatively recent time.

In contrast to the outer halo, the enhanced [Mg/Fe] ratios ($\sim 0.4$-0.5) 
frequently observed among the inner halo stars are 
compatible that the dominant fraction of the {\it inner} 
halo formed with an early accretion of a few massive progenitors 
\citep{key-20}. Alternatively, recent simulation results suggest that 
a sizable fraction of the inner halo 
may have formed in-situ \citep{key-48}. 

Further numerical simulations which investigate different formation 
mechanisms for the inner and the outer halo 
components taking into account evolution of 
both kinematics and chemical abundances will be useful to 
understand observational data as presented in this work. 

In the present study, quantitative estimation of the properties 
of the merging history such as a typical progenitor mass, 
epoch of the dominant accretion events or star formation 
history within the progenitor system remains unresolved. 
The major issue that hampers the interpretation of the 
present data is the bias against stars with $Z_{\rm max}\leqq5$ kpc
introduced in the sample selection. In order to supplement 
the lower $Z_{\rm max}$ objects, we have used abundances published 
in the literature, for which systematic differences caused 
by using the different analysis methods is present for some 
elements. The incompleteness 
of the low $Z_{\rm max}$ populations not only obscures the true 
nature of the inner halo population but may also produce
an artificial trend in the abundance ratios as a function of 
kinematics. Therefore, systematic spectroscopic surveys that are 
unbiased to particular kinematics are required for 
a more quantitative examination of the merging 
history of the MW. This should be addressed with  
advanced instruments that are capable of a multi-object 
spectroscopy down to fainter magnitudes. 
Another important task for the reconstruction of the 
MW formation is the detailed theoretical understanding 
of chemical enrichment history within the progenitor 
systems and their assembly process to form a large 
galaxy (e.g., \cite{key-22}). Our crude underlying assumption in the 
previous discussions is that the abundance 
ratios are related to the global properties of the 
progenitor systems such as mass, luminosity or 
average metallicity. These assumptions should be tested with 
theoretical modeling of these systems taking into account the 
physics of star formation, nucleosynthesis products of 
various SNe, mixing of the enriched gas or mass loss from 
evolved stars. In addition, observational estimates of 
abundance patterns for individual stars in the dwarf satellites 
should be further obtained to assess the theoretical modelings.

\section{Conclusion}
\label{sec:conc}

We present the detailed chemical abundances of 57 
metal-poor ([Fe/H]$<-1$) halo stars with $Z_{\rm max}>5$ kpc 
that are candidates of the outer halo population derived from 
homogeneous analysis of spectra obtained with Subaru/HDS.
This data set is combined with the data presented in the literature
 in which both high-resolution abundance estimates and kinematic data
are available. The resulting sample of $>$200 stars with 
[Fe/H]$<0$ are used to investigate 
systematic differences in an elemental abundance pattern 
between the presumed inner and the outer halo populations.

It is shown that the outer halo sample, conventionally 
defined as stars with $Z_{\rm max}>5$ kpc, shows lower 
[Mg/Fe] on average  by $\sim 0.1$ dex than those of the 
inner halo sample in the metallicity range of $-2<$[Fe/H]$<-1$. 
Modestly lower [Na/Fe], [Ca/Fe], [Mn/Fe], [Zn,Fe] and [Y/Fe] are 
also observed for the outer halo sample. 
Since the criterion of $Z_{\rm max}>5$ kpc preferentially 
selects the stars with a large $R_{\rm apo}$ or a large 
$V_{\rm RF}$,  the results are consistent with the earlier 
suggestions for a decrease in [$\alpha$/Fe] ratios with 
increasing $R_{\rm apo}$ or $V_{\rm RF}$ in SB02 and 
\citet{key-8}. The systematic difference in abundance 
ratios found for these elements implies that formation 
mechanisms and/or dominant building blocks between the 
inner and the outer halo are different as suggested from 
studies of kinematics and metallicity for the halo stars. 
A larger number of sample for which both high-resolution 
abundances and accurate kinematics are available will be 
required in order to perform an analysis of chemical abundance 
ratios for each bin of [Fe/H] and kinematic parameters that 
can be compared to the predictions of the standard theory 
for galaxy formation.  
 
The [Na/Fe], [Mg/Fe] and [Zn/Fe] ratios for the outer halo 
population partly overlap with those measured for bright giant 
stars in the MW satellites. On the other hand, for Ti 
and Y, neither of the inner nor outer halo populations 
overlap with the MW satellites. Whether the currently 
observed MW satellites could have contributed to the 
current stellar halo remains unclear. 
 Elaborated modeling of the star formation history of 
chemical evolution within the dwarf satellites including 
the extent of mixing of SNe ejecta as well as dynamical 
evolution of the satellites will be needed to constrain 
the detailed properties of possible building blocks of 
the MW outer halo.

We would like to thank the staff and crew at the 
Subaru observatory for their exceptional assistance on 
our observation.
We are grateful to L. Zhang, G. Zhao, T. Tsujimoto, 
T. C. Beers for useful discussion and suggestions, which 
significantly improved this paper. We also thank the 
referee, P. Nissen., for valuable comments on an earlier 
draft. This work has been 
supported in part by a Grant-in-Aid for Scientific 
Research (20340039) of the Ministry of Education, 
Culture, Sports, Science and Technology in Japan.

\appendix

\begin{table*}
\caption{Observational data }\label{obslog}
\begin{center}
\begin{tabular}{lccccccc}\hline
Star name & RA & DEC & {\it V} & [Fe/H]\footnotemark[$*$] & Exp.time & S/N\footnotemark[$\dagger$] & $V_{\rm rad}$\footnotemark[$\ddagger$]\\ 
 &(J2000) & (J2000)& (mag) & (dex) & (s) & & (km s$^{-1}$) \\\hline
 BD$+$\timeform{04D} 2621 & 12 28 44.65 & $+04$ 01 27.0 &  9.92 & $-2.41$ & 2700 &  312 & $  -41.2 \pm 0.2$\\ 
 BD$+$\timeform{13D} 2995 & 15 44 14.74 & $+12$ 56 49.7 & 11.90 & $-1.43$ & 3600 &  168 & $   45.5 \pm 0.3$ \\ 
 BD$-$\timeform{03D} 5215 & 21 28 01.31 & $-03$ 07 40.7 & 10.17 & $-1.57$ & 3600 &  369 & $ -292.8 \pm 0.2$\\ 
  BD$-$\timeform{10D} 548 & 02 46 55.72 & $-10$ 07 35.0 & 10.41 & $-1.07$ & 1200 &  281 & $  238.4 \pm 0.3$\\ 
 BD$-$\timeform{14D} 5890 & 20 56 09.15 & $-13$ 31 16.9 & 10.21 & $-2.07$ & 1200 &  359 & $  119.9 \pm 0.3$\\ 
  BD$-$\timeform{17D} 484 & 02 31 25.76 & $-16$ 59 05.1 & 10.44 & $-1.74$ & 1800 &  303 & $  236.2 \pm 0.2$\\
\end{tabular}
\end{center}
\end{table*}

\begin{longtable}{lcccccccc} 
\caption{Kinematics of the sample stars}\label{kin_table}
\hline
Star name & $U$ & $V$ & $W$ & $V_{\phi}$ & $R_{\rm apo}$ & $R_{\rm peri}$ & $Z_{\rm max}$ & $e$ \\
  & (km s$^{-1}$)& (km s$^{-1}$)& (km s$^{-1}$)& (km s$^{-1}$) & (kpc) & (kpc) & (kpc) & \\
\hline
\endfirsthead
\hline
Star name & $U$ & $V$ & $W$ & $V_{\phi}$ & $R_{\rm apo}$ & $R_{\rm peri}$ & $Z_{\rm max}$ & $e$ \\
  & (km s$^{-1}$)& (km s$^{-1}$)& (km s$^{-1}$)& (km s$^{-1}$) & (kpc) & (kpc) & (kpc) & \\
\hline
\endhead
\hline
\endfoot
\endlastfoot
 BD$+$\timeform{01D} 3070 & $   351$ & $    56$ & $  -105$ & $     275$ &  67.32 &   4.65&   17.19&    0.87 \\
 BD$+$\timeform{04D} 2466 & $  -107$ & $  -538$ & $  -259$ & $    -329$ &  54.26 &   8.47&   31.20&    0.73 \\
 BD$+$\timeform{04D} 2621 & $     2$ & $  -268$ & $  -154$ & $     -48$ &   8.47 &   2.80&    7.56&    0.53 \\
 BD$+$\timeform{09D} 2870 & $   377$ & $  -273$ & $    55$ & $     -49$ &  36.62 &   0.76&    6.48&    0.96 \\
 BD$+$\timeform{10D} 2495 & $    73$ & $   -47$ & $   275$ & $     174$ &  21.49 &   7.81&   17.30&    0.47 \\
 BD$+$\timeform{12D} 2547 & $  -299$ & $  -350$ & $  -119$ & $    -144$ &  28.79 &   2.80&    8.09&    0.82 \\
 BD$+$\timeform{13D} 2995 & $    55$ & $   -43$ & $   140$ & $     176$ &  10.03 &   6.80&    4.50&    0.20 \\
 BD$+$\timeform{29D} 2356 & $   -24$ & $   -71$ & $  -190$ & $     149$ &  10.06 &   7.81&    7.19&    0.14 \\
\hline
\end{longtable}

\begin{table*}
\caption{Measured equivalent widths\footnotemark[$*$]}\label{ews}
\begin{center}
{\tabcolsep = 1mm
\begin{tabular}{ccccccccc}
\hline
Z Elem. I/II\footnotemark[$\dagger$] &  $\lambda$\footnotemark[$\ddagger$]  &      log({\it gf})\footnotemark[$\S$]  &  Exp.\footnotemark[$\|$]  & EW\footnotemark[$\#$](BD$+$01 3070)& EW(BD$+$04 2466)& EW(BD$+$04 2621) & $\cdots$ & Ref.\footnotemark[$**$]\\
 & ({\AA})& &(eV) & ({m\AA})& ({m\AA}) & ({m\AA}) & $\cdots$ & \\ \hline
11 Na I & 5682.650  & $-0.820$ &2.10 &$10.3$ &$4.8$ &$-9.9$ & $\cdots$ & \\ 
11 Na I & 5688.219  & $-0.370$ &2.10 &$17.6$ &$-9.9$ &$4.6$ & $\cdots$ & \\
11 Na I & 6154.227  & $-1.660$ &2.10 &$-9.9$ &$-9.9$ &$-9.9$ & $\cdots$& \\

$\vdots$   &         &        &    &   &      &  $\vdots$ &   & \\
\hline
\multicolumn{9}{@{}l@{}}{\hbox to 0pt{\parbox{180mm}{\footnotesize
\par\noindent
\footnotemark[$*$]This table is fully available in an electronic form.
\par\noindent
\footnotemark[$\dagger$]Atomic number (Z), a name of the element and 
a state of inization (I/II).
\par\noindent
\footnotemark[$\ddagger$]Wavelength of a line center in units of {\AA}. 
\par\noindent
\footnotemark[$\S$]log({\it gf}) values.
\par\noindent
\footnotemark[$\|$]Excitation potential in units of electron volts.
\par\noindent
\footnotemark[$\#$]Measured equivalent widths in units of m{\AA}. 
The value of -9.9 means that the line is undetectable or not adequetly measured.
\par\noindent
\footnotemark[$**$]Reference of the atomic data.
}\hss}}
\end{tabular}
}
\end{center}
\end{table*}

\begin{longtable}{lccccccc}
 \caption{Stellar atmospheric parameters}\label{stpm}
\hline
Star name& $T_{\rm eff}$ & $\log g$ & [Fe/H] &$\xi$ & $E(B-V)$\footnotemark[$*$] & $(V-K)_{0}$\footnotemark[$\dagger$] & $T_{\rm eff}$(color) \\ 
 & (K)& & (dex)& (km s$^{-1}$)& (mag)& (mag)& (K)\\ 
\hline
\endfirsthead
\hline
Star name& $T_{\rm eff}$ & $\log g$ & [Fe/H] &$\xi$ & $E(B-V)$\footnotemark[$*$] & $(V-K)_{0}$\footnotemark[$\dagger$] & $T_{\rm eff}$(color) \\ 
 & (K)& & (dex)& (km s$^{-1}$)& (mag)& (mag)& (K)\\ 
\hline
\endhead
\hline
\endfoot
\endlastfoot
BD$+$\timeform{01D} 3070 &  5077$\pm$35 & 3.01 & $-1.63$ & 1.04 &  0.06 &  1.83 &  5308 \\ 
BD$+$\timeform{04D} 2466 &  5065$\pm$93 & 1.82 & $-2.10$ & 1.82 &  0.04 &  1.87 &  5265 \\ 
BD$+$\timeform{04D} 2621 &  4453$\pm$18 & 0.90 & $-2.68$ & 1.92 &  0.02 &  2.31 &  4760 \\ 
BD$+$\timeform{09D} 2870 &  4285$\pm$18 & 0.29 & $-2.70$ & 1.92 &  0.03 &  2.44 &  4620 \\ 
BD$+$\timeform{10D} 2495 &  4819$\pm$47 & 1.92 & $-2.17$ & 1.70 &  0.02 &  2.09 &  5003 \\ 
BD$+$\timeform{12D} 2547 &  4494$\pm$27 & 1.29 & $-2.10$ & 1.76 &  0.03 &  2.38 &  4673 \\ 
BD$+$\timeform{13D} 2995 &  5650$\pm$54 & 4.90 & $-1.07$ & 0.05 &  0.03 &  1.61 &  5561 \\ 
BD$+$\timeform{29D} 2356 &  4884$\pm$22 & 2.49 & $-1.44$ & 1.56 &  0.01 &  2.42 &  4637 \\
\end{longtable}

\begin{table*}
\caption{Comparison of atmospheric parameters derived in this work and those derived in literature}\label{comp_stpm_table}
\begin{center}
\begin{tabular}{lcccccccccccc}\hline
Name & \multicolumn{3}{c}{$T_{\rm eff}$ (K)} & \multicolumn{3}{c}{$\log g$} & \multicolumn{3}{c}{[Fe/H] (dex)} & \multicolumn{3}{c}{$\xi$ (km s$^{-1}$)}\\
 & TW & SB02 & $\Delta$\footnotemark[$*$] &TW &  SB02 & $\Delta$  &  TW & SB02 &$\Delta$  & TW & SB02 &$\Delta$ \\
\hline   
   G 15-13 & 5225 & 5082 & $ 143$ &  5.00 &  4.61 & $ 0.39$ & $-1.50$ & $-1.70$ & $ 0.20$ &  0.05 &  0.00 & $ 0.05$ \\
  G 165-39 & 6162 & 6118 & $  44$ &  3.62 &  3.53 & $ 0.09$ & $-2.20$ & $-2.17$ & $-0.03$ &  1.57 &  1.45 & $ 0.12$ \\
  G 166-37 & 5355 & 5350 & $   5$ &  4.95 &  4.71 & $ 0.24$ & $-1.31$ & $-1.39$ & $ 0.08$ &  0.05 &  0.00 & $ 0.05$ \\
  G 188-30 & 5326 & 5141 & $ 185$ &  5.00 &  4.44 & $ 0.56$ & $-1.64$ & $-1.89$ & $ 0.25$ &  0.05 &  0.31 & $-0.26$ \\
  G 238-30 & 5299 & 5383 & $ -84$ &  3.39 &  3.43 & $-0.04$ & $-3.77$ & $-3.60$ & $-0.17$ &  1.19 &  1.19 & $ 0.00$ \\
   G 25-24 & 5594 & 5733 & $-139$ &  4.07 &  3.98 & $ 0.09$ & $-2.06$ & $-1.88$ & $-0.18$ &  1.02 &  1.21 & $-0.19$ \\
\end{tabular}
\end{center}
\end{table*}

\begin{longtable}{lcccccccccc}
\caption{Derived Na, Mg, Si, and Ca abundances }\label{tab:ab1}
\hline 
Star name& [Na I/H]& $N$ & [Mg I/H]& $N$ & [Si I/H]& $N$ &[Si II/H]& $N$ & [Ca I/H]& $N$  \\ 
\hline 
\endfirsthead 
\hline 
Star name& [Na I/H]& $N$ & [Mg I/H]& $N$ & [Si I/H]& $N$ &[Si II/H]& $N$ & [Ca I/H]& $N$  \\ 
\hline 
\endhead 
\hline 
\endfoot 
\hline 
\endlastfoot 
\hline 
BD$+$\timeform{01D} 3070 & $-1.88\pm0.09$& 2 & $-1.26\pm0.03$ &  3 & $-1.38\pm0.09$ & 1 & - & - & $-1.39\pm0.03$ & 13 \\ 
BD$+$\timeform{04D} 2466 & $-2.14\pm0.08$& 1 & $-1.66\pm0.08$ &  2 & $-1.52\pm0.08$ & 1 & - & - & $-1.63\pm0.04$ & 10 \\ 
BD$+$\timeform{04D} 2621 & $-2.89\pm0.06$& 1 & $-2.09\pm0.08$ &  4 & - & - & - & - & $-2.29\pm0.02$ & 8 \\ 
BD$+$\timeform{09D} 2870 & $-2.76\pm0.08$& 1 & $-1.96\pm0.16$ &  5 & - & - & - & - & $-2.28\pm0.03$ & 11 \\ 
BD$+$\timeform{10D} 2495 & $-2.33\pm0.08$& 2 & $-1.70\pm0.06$ &  4 & $-1.82\pm0.07$ & 1 & - & - & $-1.88\pm0.03$ & 14 \\ 
BD$+$\timeform{12D} 2547 & $-2.39\pm0.08$& 2 & $-1.61\pm0.08$ &  1 & $-1.68\pm0.08$ & 1 & - & - & $-1.80\pm0.03$ & 9 \\ 
BD$+$\timeform{13D} 2995 & $-1.21\pm0.06$& 2 & $-0.52\pm0.06$ &  2 & - & - & - & - & $-0.87\pm0.03$ & 10 \\ 
BD$+$\timeform{29D} 2356 & $-1.70\pm0.06$& 3 & $-1.05\pm0.10$ &  1 & $-1.16\pm0.10$ & 2 & - & - & $-1.22\pm0.04$ & 9 \\
\hline 
\end{longtable}

\begin{longtable}{lcccccccccc}
\caption{Derived Ti, Cr and Mn abundances }\label{tab:ab2}
\hline 
Star name& [Ti I/H]& $N$ & [Ti II/H]& $N$ & [Cr I/H]& $N$&[Cr II/H]& $N$ & [Mn I/H]& $N$ \\ 
\hline 
\endfirsthead 
\hline 
Star name& [Ti I/H]& $N$ & [Ti II/H]& $N$ & [Cr I/H]& $N$&[Cr II/H]& $N$ & [Mn I/H]& $N$ \\ 
\hline 
\endhead 
\hline 
\endfoot 
\hline 
\endlastfoot 
\hline 
BD$+$\timeform{01D} 3070 & $-1.46\pm0.06$ &11 & $-1.19\pm0.04$ & 7 & $-1.82\pm0.03$ & 10 & $-1.52\pm0.05$  & 5 & $-2.32\pm0.09$ & 2 \\ 
BD$+$\timeform{04D} 2466 & $-1.94\pm0.08$ &1 & $-1.62\pm0.08$ & 2 & $-2.28\pm0.03$ & 5 & $-1.97\pm0.08$  & 1 & $-2.66\pm0.08$ & 1 \\ 
BD$+$\timeform{04D} 2621 & $-2.59\pm0.03$ &7 & $-2.28\pm0.04$ & 4 & $-3.02\pm0.02$ & 5 & $-2.59\pm0.05$  & 1 & $-3.18\pm0.02$ & 3 \\ 
BD$+$\timeform{09D} 2870 & $-2.66\pm0.02$ &10 & $-2.43\pm0.04$ & 8 & $-3.08\pm0.02$ & 7 & $-2.70\pm0.08$  & 2 & $-3.39\pm0.08$ & 1 \\ 
BD$+$\timeform{10D} 2495 & $-2.07\pm0.02$ &7 & $-1.84\pm0.03$ & 6 & $-2.44\pm0.02$ & 9 & $-2.04\pm0.06$  & 4 & $-2.75\pm0.02$ & 3 \\ 
BD$+$\timeform{12D} 2547 & $-2.07\pm0.03$ &8 & $-1.68\pm0.05$ & 5 & $-2.37\pm0.02$ & 7 & $-1.99\pm0.04$  & 3 & $-2.66\pm0.08$ & 2 \\ 
BD$+$\timeform{13D} 2995 & $-0.82\pm0.03$ &7 & $-0.47\pm0.05$ & 3 & $-1.18\pm0.04$ & 6 & $-0.96\pm0.06$  & 2 & $-1.60\pm0.03$ & 3 \\ 
BD$+$\timeform{29D} 2356 & $-1.23\pm0.06$ &10 & $-0.92\pm0.05$ & 4 & $-1.64\pm0.02$ & 6 & $-1.40\pm0.10$  & 2 & $-2.02\pm0.03$ & 3 \\
\hline 
\end{longtable} 

{\tabcolsep = 1mm
\begin{longtable}{lcccccccccccc}
\caption{Derived Fe, Ni, Zn, Y and Ba abundances }\label{tab:ab3}
\hline 
Star name& [Fe I/H]& $N$ & [Fe II/H]& $N$ & [Ni I/H]& $N$ &[Zn I/H]& $N$ & [Y II/H]& $N$ & [Ba II/H] & $N$ \\ 
\hline 
\endfirsthead 
\hline 
Star name& [Fe I/H]& $N$ & [Fe II/H]& $N$ & [Ni I/H]& $N$ &[Zn I/H]& $N$ & [Y II/H]& $N$ & [Ba II/H] & $N$ \\ 
\hline 
\endhead 
\hline 
\endfoot 
\hline 
\endlastfoot 
\hline 
BD$+$\timeform{01D} 3070 & $-1.63\pm0.01$ & 63 & $-1.63\pm0.04$ & 9 & $-1.71\pm0.05$ & 11 & $-1.74\pm0.09$  & 2 & $-1.82\pm0.09$ & 2 & $-1.76\pm0.09$ & 3 \\ 
BD$+$\timeform{04D} 2466 & $-2.10\pm0.02$ & 32 & $-2.10\pm0.06$ & 5 & $-1.96\pm0.08$ & 2 & $-2.12\pm0.09$  & 1 & $-1.56\pm0.08$ & 2 & $-0.75\pm0.08$ & 1 \\ 
BD$+$\timeform{04D} 2621 & $-2.69\pm0.01$ & 49 & $-2.67\pm0.03$ & 5 & $-2.68\pm0.06$ & 4 & $-2.54\pm0.07$  & 1 & $-3.16\pm0.06$ & 1 & $-3.74\pm0.05$ & 4 \\ 
BD$+$\timeform{09D} 2870 & $-2.69\pm0.01$ & 53 & $-2.70\pm0.03$ & 7 & $-2.75\pm0.06$ & 8 & $-2.75\pm0.09$  & 2 & $-3.25\pm0.08$ & 2 & $-3.67\pm0.08$ & 2 \\ 
BD$+$\timeform{10D} 2495 & $-2.17\pm0.01$ & 66 & $-2.17\pm0.04$ & 9 & $-2.18\pm0.04$ & 9 & $-2.24\pm0.08$  & 2 & $-2.40\pm0.08$ & 2 & $-2.43\pm0.05$ & 4 \\ 
BD$+$\timeform{12D} 2547 & $-2.11\pm0.01$ & 49 & $-2.06\pm0.06$ & 5 & $-2.20\pm0.05$ & 11 & $-2.13\pm0.09$  & 2 & $-2.26\pm0.08$ & 2 & $-2.13\pm0.08$ & 1 \\ 
BD$+$\timeform{13D} 2995 & $-1.07\pm0.01$ & 26 & $-1.08\pm0.02$ & 6 & $-1.12\pm0.06$ & 10 & $-1.08\pm0.07$  & 2 & $-0.71\pm0.06$ & 1 & $-0.71\pm0.06$ & 2 \\ 
BD$+$\timeform{29D} 2356 & $-1.44\pm0.02$ & 52 & $-1.41\pm0.06$ & 5 & $-1.48\pm0.07$ & 9 & $-1.52\pm0.11$  & 2 & $-1.38\pm0.10$ & 1 & $-1.30\pm0.10$ & 1 \\
\hline 
\end{longtable} 
}

\begin{longtable}{lcccccc}
\caption{Fe-normalized abundance ratios }\label{tab:ab4}
\hline 
Star name& [Fe/H]& [Na/Fe]& [Mg/Fe]&  [Si/Fe]&[Ca/Fe]& [Ti/Fe] \\ 
\hline 
\endfirsthead 
\hline 
Star name& [Fe/H]& [Na/Fe]& [Mg/Fe]&  [Si/Fe]&[Ca/Fe]& [Ti/Fe] \\ 
\hline 
\endhead 
\hline 
\endfoot 
\hline 
\endlastfoot 
\hline 
BD$+$\timeform{01D} 3070 & $-1.63\pm0.01$ & $-0.25\pm0.09$ & $ 0.37\pm0.03$ & $ 0.25\pm0.09$ & $ 0.24\pm0.03$ & $ 0.35\pm0.04$   \\ 
BD$+$\timeform{04D} 2466 & $-2.10\pm0.02$ & $-0.04\pm0.09$ & $ 0.44\pm0.08$ & $ 0.58\pm0.08$ & $ 0.48\pm0.04$ & $ 0.33\pm0.06$   \\ 
BD$+$\timeform{04D} 2621 & $-2.68\pm0.01$ & $-0.21\pm0.06$ & $ 0.59\pm0.09$ & - & $ 0.40\pm0.03$ & $ 0.21\pm0.03$   \\ 
BD$+$\timeform{09D} 2870 & $-2.70\pm0.01$ & $-0.07\pm0.08$ & $ 0.74\pm0.16$ & - & $ 0.42\pm0.03$ & $ 0.11\pm0.02$   \\ 
BD$+$\timeform{10D} 2495 & $-2.17\pm0.01$ & $-0.16\pm0.08$ & $ 0.47\pm0.06$ & $ 0.35\pm0.07$ & $ 0.29\pm0.03$ & $ 0.19\pm0.02$   \\ 
BD$+$\timeform{12D} 2547 & $-2.10\pm0.01$ & $-0.29\pm0.08$ & $ 0.50\pm0.08$ & $ 0.42\pm0.08$ & $ 0.30\pm0.03$ & $ 0.16\pm0.03$   \\ 
BD$+$\timeform{13D} 2995 & $-1.07\pm0.01$ & $-0.13\pm0.06$ & $ 0.55\pm0.06$ & - & $ 0.20\pm0.03$ & $ 0.35\pm0.03$   \\ 
BD$+$\timeform{29D} 2356 & $-1.44\pm0.02$ & $-0.26\pm0.06$ & $ 0.39\pm0.10$ & $ 0.28\pm0.10$ & $ 0.22\pm0.04$ & $ 0.39\pm0.04$   \\
\hline 
\end{longtable}

\begin{longtable}{lcccccc}
\caption{Fe-normalized abundance ratios }\label{tab:ab5}
\hline 
Star name& [Cr/Fe] & [Mn/Fe]& [Ni/Fe]&[Zn/Fe]&[Y/Fe] & [Ba/Fe] \\ 
\hline 
\endfirsthead 
\hline 
Star name& [Cr/Fe] & [Mn/Fe]& [Ni/Fe]&[Zn/Fe]&[Y/Fe] & [Ba/Fe] \\ 
\hline 
\endhead 
\hline 
\endfoot 
\hline 
\endlastfoot 
\hline 
BD$+$\timeform{01D} 3070 & $-0.12\pm0.03$ & $ 0.09\pm0.09$  &$-0.08\pm0.05$ & $-0.11\pm0.10$ & $-0.19\pm0.09$ & $-0.13\pm0.09$ \\ 
BD$+$\timeform{04D} 2466 & $-0.14\pm0.03$ & $ 0.08\pm0.08$  &$ 0.15\pm0.08$ & $-0.01\pm0.09$ & $ 0.54\pm0.09$ & $ 1.36\pm0.09$ \\ 
BD$+$\timeform{04D} 2621 & $-0.27\pm0.02$ & $ 0.03\pm0.03$  &$-0.00\pm0.06$ & $ 0.15\pm0.07$ & $-0.48\pm0.06$ & $-1.06\pm0.05$ \\ 
BD$+$\timeform{09D} 2870 & $-0.37\pm0.02$ & $ 0.08\pm0.08$  &$-0.05\pm0.06$ & $-0.06\pm0.09$ & $-0.56\pm0.08$ & $-0.98\pm0.08$ \\ 
BD$+$\timeform{10D} 2495 & $-0.22\pm0.02$ & $ 0.02\pm0.02$  &$-0.01\pm0.04$ & $-0.07\pm0.09$ & $-0.23\pm0.08$ & $-0.26\pm0.05$ \\ 
BD$+$\timeform{12D} 2547 & $-0.19\pm0.02$ & $ 0.08\pm0.08$  &$-0.09\pm0.05$ & $-0.02\pm0.09$ & $-0.15\pm0.08$ & $-0.03\pm0.08$ \\ 
BD$+$\timeform{13D} 2995 & $-0.03\pm0.03$ & $ 0.04\pm0.04$  &$-0.04\pm0.06$ & $-0.00\pm0.07$ & $ 0.37\pm0.06$ & $ 0.36\pm0.06$ \\ 
BD$+$\timeform{29D} 2356 & $-0.19\pm0.03$ & $ 0.04\pm0.04$  &$-0.04\pm0.07$ & $-0.09\pm0.11$ & $ 0.06\pm0.10$ & $ 0.14\pm0.10$ \\
\hline 
\end{longtable}

\begin{table*}
\caption{Averages and scatters of abundances for the inner/outer halo samples.}\label{abaverage}
\begin{center}
\begin{tabular}{lccccccccccccc}\hline
Element & $Z_{\rm max}$  & \multicolumn{3}{c}{$-4<$[Fe/H]$\leqq-3$\footnotemark[$*$]} & \multicolumn{3}{c}{$-3<$[Fe/H]$\leqq-2$} & \multicolumn{3}{c}{$-2<$[Fe/H]$\leqq-1$}& \multicolumn{3}{c}{$-1<$[Fe/H]$\leqq0$} \\
 & (kpc) & $\mu$\footnotemark[$\dagger$] & $\frac{\sigma}{\sqrt{\mathstrut N}}$\footnotemark[$\ddagger$] & $N$\footnotemark[\S] & $\mu$ & $\frac{\sigma}{\sqrt{\mathstrut N}}$ & $N$ & $\mu$ & $\frac{\sigma}{\sqrt{\mathstrut N}}$ & $N$ & $\mu$ & $\frac{\sigma}{\sqrt{\mathstrut N}}$ & $N$ \\
\hline
 Na & $\leqq$5  & -& -&   0 & $  0.15$ & $  0.14$ &   7 & $ -0.10$ & $  0.02$ &  52 & $  0.07$ & $  0.02$ &  65\\
 & $>$5 & -& -&   0 & $ -0.23$ & $  0.04$ &  18 & $ -0.31$ & $  0.04$ &  39 & $ -0.26$ & $  0.03$ &  2\\
 Mg & $\leqq$5  & -& -&   1 & $  0.44$ & $  0.02$ &  16 & $  0.42$ & $  0.02$ &  51 & $  0.39$ & $  0.01$ &  65\\
 & $>$5 & $  0.66$ & $  0.05$ &   7 & $  0.49$ & $  0.03$ &  29 & $  0.31$ & $  0.02$ &  40 & $  0.31$ & $  0.25$ &  2\\
 Si & $\leqq$5  & -& -&   0 & $  0.46$ & $  0.04$ &   6 & $  0.30$ & $  0.02$ &  46 & $  0.24$ & $  0.01$ &  65\\
 & $>$5 & -& -&   0 & $  0.43$ & $  0.03$ &   7 & $  0.25$ & $  0.02$ &  26 & $  0.06$ & $  0.08$ &  2\\
\end{tabular}
\end{center}
\end{table*}

\begin{table}
\caption{Dependence of P$_{\mathrm{KS}}$ on the adopted halo boundary}\label{ks}
\begin{center}
\begin{tabular}{lcccc} \hline
 Boundary & \multicolumn{4}{c}{P$_{\rm KS}$\footnotemark[$*$]}\\ 
$Z_{\rm max}$ (kpc) & [Mg/Fe]& [Si/Fe]&[Ca/Fe]& [Ti/Fe]\\ \hline
3 &0.13$\times 10^{-3}$ &  0.70$\times 10^{-1}$ &  0.36$\times 10^{-3}$&  0.79\\
5 & 0.30$\times 10^{-4}$ &  0.39 &  0.60$\times 10^{-3}$ &  0.15\\
10 &  0.45$\times 10^{-2}$&  0.90$\times 10^{-1}$ &  0.11$\times 10^{-1}$ &  0.57\\
\hline
\multicolumn{5}{@{}l@{}}{\hbox to 0pt{\parbox{85mm}{\footnotesize
\par\noindent
\footnotemark[$*$] Kolmogorov-Smirnoff probabilities for the inner/outer halo
division on the boundary $Z_{\rm max}$ value.
}\hss}}
\end{tabular}
\end{center}
\end{table}

\clearpage

\begin{thebibliography}{100}
\bibitem[Alonso et al.(1996)]{key-32}
Alonso, A., Arribas, S., \& Martinez-Roger, C. 1996, \aap, 313, 873
\bibitem[Alonso et al.(1999)]{key-33}
Alonso, A., Arribas, S., \& Martinez-Roger, C. 1999, \aaps, 140, 261
\bibitem[An et al.(2009)]{key-91}
An, D. \etal\ 2009, arXiv:0907.1082
\bibitem[Andrievsky et al.(2007)]{key-38}
Andrievsky, S. M., Spite, M., Korotin, S. A., Spite, F., Bonifacio, P., Cayrel, R., Hill, V., \& Francois, P. 2007, \aap, 464, 1081
\bibitem[Aoki et al.(2005)]{key-10}
Aoki, W. \etal\ 2005, {\apj}, 632, 611
\bibitem[Aoki et al.(2009a)]{key-45}
Aoki, W. \etal\ 2009a, \aap, 502, 569
\bibitem[Aoki et al.(2009b)]{key-79}
Aoki, W., Barklem, P. S., Beers, T. C., Christlieb, N., Inoue, S., Garc{\'i}a P{\'e}rez, A. E., Norris, J. E., Carollo, D. 2009b, \apj, 698, 1803
\bibitem[Baba et al.(2002)]{key-83}
Baba, H., \etal\ 2002, in Astronomical Data Analysis Software and Systems XI, APS Conference Proceedings Vol. 281, ed. D. A. Bohlender, D. Durand, \& T. H. Handley (San Francisco: Astronomical Society of the Pacific), 298 
\bibitem[Barklem et al.(2005)]{key-84}
Barklem, P. S., \etal\ 2005, \aap, 439, 129
\bibitem[Baumuller et al.(1998)]{key-39}
Baumuller, D., Butler, K., \& Gehren, T. 1998, \aap, 338, 637
\bibitem[Beers et al.(2000)]{key-5}
Beers, T. C., Chiba, M., Yoshii, Y., Platais, I., Hanson, R. B., Fuchs, B., \& Rossi, S. 2000, {\aj}, 119, 2866
\bibitem[Bell et al.(2008)]{key-64}
Bell, E. F. \etal\ 2008, \apj, 680, 295
\bibitem[Belokurov et al.(2006)]{key-60}
Belokurov, V. \etal\ 2006, \apj, 642, L137
\bibitem[Belokurov et al.(2007)]{key-41}
Belokurov, V. \etal\ 2007, \apj, 654, 897
\bibitem[Brown et al.(2008)]{key-61}
Brown, T. M. \etal\ 2008, \apj, 685, L121
\bibitem[Bullock \& Johnston(2005)]{key-25}
Bullock, J. S. \& Johnston, K. V. 2005, {\apj}, 635, 931
\bibitem[Carney et al.(1994)]{key-6}
Carney, B. W., Latham, D. W., Laird, J. B. \& Aguilar, L. A. 1994, {\aj}, 107, 2240
\bibitem[Carney et al.(2003)]{key-29}
Carney, B. W., Latham, D. W., Stefanik, R. P., Laird, J. B. \& Morse, J. A. 2003, {\aj}, 125, 293
\bibitem[Carollo et al.(2007)]{key-35}
Carollo, D. \etal\ 2007, Nature, 450, 1020
\bibitem[Carollo et al. (2009)]{key-90}
Carollo, D. \etal\ 2009, arXiv:0909.3019
\bibitem[Castelli \& Kurucz(2003)]{key-77}
Castelli, F. \& Kurucz, R. L. 2003, IAUS, 210P, A20
\bibitem[Cayrel(1988)]{key-31}
Cayrel, R. 1988, in The Impact of Very High S/N Spectroscopy on Stellar Physics, ed. G. Cayrel de Strobel, \& M. Spite (Dordrect: Kluwer), 345
\bibitem[Chiba \& Beers(2000)]{key-4}
Chiba, M. \& Beers, T. C. 2000, {\aj}, 119, 2843
\bibitem[De Lucia \& Helmi(2008)]{key-58}
De Lucia, G. \& Helmi, A. 2008, \mnras, 391, 14
\bibitem[Dettbarn et al.(2007)]{key-52}
Dettbarn, C., Fuchs, B., Flynn, C., \& Williams, M. \aap, 474, 857
\bibitem[Diemand et al.(2007)]{key-59}
Diemand, J., Kuhlen, M. \& Madau, P. 2007, \apj, 667, 859
\bibitem[Font et al.(2006a)]{key-20}
Font, A. S., Johnston, K. V., Bullock, J. S., \& Robertson, B. E. 2006a, \apj, 638, 585
\bibitem[Font et al.(2006b)]{key-55}
Font, A. S., Johnston, K. V., Bullock, J. S., \& Robertson, B. E. 2006b, \apj, 646, 886 
\bibitem[Francois et al.(2007)]{key-1}
Francois, P., Depagne, E. \& Hill, V., et al. 2007, {\aap}, 476, 935
\bibitem[Frebel et al.(2009)]{key-44}
Frebel, A., Simon, J. D., Geha, M. \& Willman, B. 2009, arXiv:0902.2395
\bibitem[Freeman \& Bland-Hawthorn(2002)]{key-67}
Freeman, K. C. \& Bland-Hawthorn, J. 2002, \araa, 40, 487
\bibitem[Fulbright(2002)]{key-8}
Fulbright, J.P. 2002, {\aj}, 123, 404
\bibitem[Geisler et al.(2005)]{key-82}
Geisler, D., Smith, V. V., Wallerstein, G., Gonzalez, G., \& Charbonnel, C. 2005, \aj, 129, 1428
\bibitem[Geisler et al.(2007)]{key-46}
Geisler, D., Wallerstein, G., Smith, V. V. \& Casetti-Dinescu, D. I. 2007, 
PASP, 119, 939
\bibitem[Gratton et al.(2003a)]{key-72}
Gratton, R. G., Carretta, E., Claudi, R., Lucatello, S. \& Barberi, M. 2003, {\aap}, 404, 187 (G03)
\bibitem[Gratton et al.(2003b)]{key-9}
Gratton, R. G., Carretta, E., Desidera, S., Lucatello, S., Mazzei, P. \& Barbieri, M. 2003, {\aap}, 406, 131 
\bibitem[Grevesse et al.(1996)]{key-34}
Grevesse, N., Noels, A. \& Sauval, A. J. 1996, ASPC, 99, 117
\bibitem[Hanson et al.(1998)]{key-69}
Hanson, R. B., Sneden, C., Kraft, R. P. \& Fulbright, J. 1998, \aj, 116, 1286
\bibitem[Helmi \& White(1999)]{key-66}
Helmi, A. \& White, S. D. M. 1999, \mnras, 307, 495
\bibitem[Helmi et al.(1999)]{key-26}
Helmi, A., White, S. D. M., de Zeeuw, T. \& Zhao, H. 1999, {\nat}, 402, 53
\bibitem[Ibata et al.(1994)]{key-18}
Ibata, R. A., Gilmore, G. \& Irwin, M. J. 1994, Nature, 370, 194
\bibitem[Ivans et al.(2003)]{key-16}
Ivans, I. I., Sneden, C., James, C. R., Preston, G. W., Fulbright, J. P., Hoflich, P. A., Carney, B. W. \& Wheeler, J. G. 2003, {\apj}, 592, 906
\bibitem[Ivezic et al.(2008)]{key-24}
Ivezic, Z. \etal\ 2008, {\apj}, 684, 287 
\bibitem[Johnston et al.(2008)]{key-22}
Johnston, K. V., Bullock, J. S., Sharma, S., Font, A., Robertson, B. E. \& Leitner, S. N. 2008, {\apj}, 689, 936
\bibitem[Jorissen et al.(2005)]{key-78}
Jorissen, A., Za{\v c}s, L., Ubry, S., Lindgren, H. and Musaev, F. A. 2005, \aap, 441, 1135
\bibitem[Juric et al.(2008)]{key-23}
Juric, M. \etal\ 2008, {\apj}, 673, 864
\bibitem[Kazantzidis et al.(2009)]{key-49}
Kazantzidis, S., Zentner, A. R., Kravtsov, A. V., Bullock, J. S. \& Debattista, V. P. 2009, \apj, 700, 1896
\bibitem[Kepley et al.(2007)]{key-27}
Kepley, A. A. \etal\ 2007, {\aj}, 134, 1579
\bibitem[Kirby et al.(2008)]{key-43}
Kirby, E. N., Simon, J. D., Geha, M., Guhathakurta, P. and Frebel, A. 2008, \apj, 685, L43
\bibitem[Klement et al.(2009)]{key-53}
Klement, R. \etal\ 2009, \apj, 698, 865
\bibitem[Koch et al.(2008a)]{key-13}
Koch, A., Grebel, E. K., Gilmore, G. F., Wyse, R. F. G., Kleyna, J. T., Harbeck, D. R., Wilkinson, M. I. \& Evans, N. W. 2008a, {\aj}, 135, 1580
\bibitem[Koch et al.(2008b)]{key-14}
Koch, A., McWilliam, A., Grebel, E. K., Zucker, D. B. \& Belokurov, V. 2008b, {\apj}, 688, L13	
\bibitem[Kupka et al.(2000)]{key-85}	
Kupka, F. G., Ryabchikova, T. A., Piskunov, N. E., Stempels, H. C. \& Weiss, W. W. 2000, Baltic Astronomy, 9, 590
\bibitem[Kurucz \& Bell(1995)]{key-88}
Kurucz, R. L., \& Bell, B. 1995, Kurucz CD-ROM 23, Atomic Line Data (Cambridge: SAO)
\bibitem[Lai et al.(2008)]{key-50}
Lai, D. K., Bolte, M., Johnson, J. A., Lucatello, S., Heger, A. \& Woosley, S. E. 2008, \apj, 681, 1524
\bibitem[Lanfranchi \& Matteucci(2003)]{key-73}
Lanfranchi, G. A. \& Matteucci, F. 2003, \mnras, 345, 71
\bibitem[Latham et al.(2002)]{key-30}
Latham, D. W., Stefanik, R. P., Torres, G., Davis, R. J., Mazeh, T., Carney, B. W., Laird, J. B. \& Morse, J. A. 2002, {\aj}, 124, 1144
\bibitem[Majewski et al.(2003)]{key-21} 
Majewski, S. R., Skrutskie, M. F., Weinberg, M. D. \& Ostheimer, J. C. 2003, {\apj}, 599, 1082
\bibitem[Martinez-Delgado et al.(2008)]{key-63}
Martinez-Delgado, D., Penarrubia, J., Gabany, R. J., Trujillo, I., Majewski, S. R. \& Pohlen, M. 2008, \apj, 689, 184
\bibitem[Matteucci \& Greggio(1986)]{key-71}
Matteucci, F., \& Greggio, L. 1986, \aap, 154, 279
\bibitem[McWilliam(1998)]{key-11}
McWilliam, A. 1998, {\aj}, 1998, 115, 1640
\bibitem[Morrison et al.(2009)]{key-56}
Morrison, H. L. \etal\ 2009, \apj, 694, 130
\bibitem[Nissen \& Schuster(1997)]{key-68}
Nissen, P. E., \& Schuster, W. J. 1997, \aap, 326, 751
\bibitem[Noguchi et al.(2002)]{key-75}
Noguchi, K. \etal\ 2002, \pasj, 54, 855
\bibitem[Norris(1994)]{key-80}
Norris, J. E. 1994, \apj, 431, 645
\bibitem[Richardson et al.(2009)]{key-62}
Richardson, J. C. \etal\ 2009, \mnras, 396, 1842
\bibitem[Robertson et al.(2005)]{key-19}
Robertson, B., Bullock, J. S., Font, A. S., Johnston, V. \& Hernquist, L. 2005, {\apj}, 632, 872
\bibitem[Roederer(2008)]{key-36}
Roederer, I. U. 2008, {\aj}, 137, 272
\bibitem[Ryan \& Norris(1991)]{key-37}
Ryan, S. G. \& Norris, J. E. 1991, {\aj}, 101, 1835
\bibitem[Ryan, Norris, \& Beers(1996)]{key-87}
Ryan, S. G., Norris, J. E. \& Beers, T. C. 1996, \apj, 471, 254
\bibitem[Schlaufman et al.(2009)]{key-89}
Schlaufman, K. C. \etal\ 2009, \apj, 703, 2177
\bibitem[Searle \& Zinn(1978)]{key-47}
Searle, L. \& Zinn, R. 1978, ApJ, 225, 357
\bibitem[Shetrone et al.(2001)]{key-2}
Shetrone, M. D., Cote, P. \& Sargent, W. L. W. 2001, {\apj}, 548, 529
\bibitem[Shetrone et al.(2003)]{key-12}
Shetrone, M. D., Venn, K. A., Tolstoy, E., Primas, F., Hill, V. \& Kaufer, A. 2003, {\apj}, 125, 684
\bibitem[Skrutskie et al.(2006)]{key-57}
Skrutskie, M. F. \etal\ 2006, AJ, 131, 1163
\bibitem[Sobeck et al.(2007)]{key-51}
Sobeck, J. S., Lawler, J. E. \& Sneden, C. 2007, \apj, 667, 1267
\bibitem[Sommer-Larsen \& Zhen(1990)]{key-81}
Sommer-Larsen, J. \& Zhen, C., 1990, \mnras, 242, 10
\bibitem[Starkenburg et al.(2009)]{key-65}
Starkenburg, E. \etal\ 2009, \apj, 698, 567
\bibitem[Stephens(1999)]{key-54}
Stephens, A. 1999, \aj, 117, 1771
\bibitem[Stephens \& Boesgaard(2002)]{key-7}
Stephens, A. \& Boesgaard, A. M. 2002, {\aj}, 123, 1647 (SB02)
\bibitem[Takeda et al.(2003)]{key-76}
Takeda, Y., Zhao, G., Takada-Hidai, M., Chen, Y.-Q., Saito, Y., \& Zhang, H.-W. 2003, Chin. J. Astron. Astrophys., 3, 316 
\bibitem[Tolstoy et al.(2003)]{key-74}
Tolstoy, E., Venn, K. A., Shetrone, M., Hill, V., Kaufer, A., \& Szeifert, T. 2003, \aj, 125, 707
\bibitem[Umeda \& Nomoto(2002)]{key-17}
Umeda, H. \& Nomoto, K. 2002, {\apj}, 565, 385
\bibitem[Uns\"{o}ld(1955)]{key-86}
Uns\"{o}ld, A.\ 1955, Physik der Sternatmospharen, MIT besonderer Berucksichtigung der Sonne (Berlin, Springer), 2. Aufl.
\bibitem[Venn et al.(2004)]{key-15}
Venn, K. A., Irwin, M., Shetrone, M. D., Tout, C. A., Hill, V. \& Tolstoy, E. 2004, {\aj}, 128, 1177
\bibitem[Willman et al.(2005)]{key-42}
Willman, B. \etal\ 2005, AJ, 129, 2692
\bibitem[Woosley \& Weaver(1995)]{key-70}
Woosley, W. E., \& Weaver, T. A. 1995, \apjs, 101, 181
\bibitem[Zhang et al.(2009)]{key-3}
Zhang, L., Ishigaki, M., Aoki, W., Zhao, G. \& Chiba, M. 2009, \apj, 706, 1095
\bibitem[Zolotov et al.(2009)]{key-48}
Zolotov, A., Willman, B., Brooks, A. M., Governato, F., Brook, C. B., 
Hogg, D. W., Quinn, T. \& Stinson, G. 2009, \apj, 702, 1058
\end{thebibliography}
\end{document}